\documentstyle[pra,aps,eqsecnum]{revtex}

\def\ut#1{\lower1.2ex\hbox{$\mathchar"3218$}\mkern -14mu%
          \hbox to 2ex{\hss$#1$\hss}}

\begin{document}
\draft

\title{Condensate fluctuations in finite Bose-Einstein condensates at finite
temperature}
\author{Robert Graham}
\address{Fachbereich Physik, Universit\"at-Gesamthochschule Essen\\
45117 Essen\\ Germany}
\maketitle

\begin{abstract}
A  Langevin equation for the complex amplitude of a single-mode
Bose-Einstein condensate is derived. The equation is first formulated phenomenologically, defining three transport parameters. It is then also derived microscopically. Expressions for the transport parameters in the form of Green-Kubo formulas are thereby derived and evaluated for simple trap geometries,
a cubic box with cyclic boundary conditions and an isotropic parabolic trap.
The number fluctuations in the condensate,
their correlation time $\tau_c$, and the temperature-dependent collapse-time
of the order parameter as well as its phase-diffusion coefficient are calculated,  
\end{abstract}

\vskip2pc

\section{Introduction}\label{sec:1}

Bose-Einstein condensation in a weakly interacting Bose-gas in three
dimensions in the thermodynamic limit of an  infinitely extended system is a
second order phase transition in which an order parameter, the macroscopic
wave-function, appears spontaneously with a fixed but arbitrary phase,
turning the global $U(1)$-gauge-symmetry connected with particle-number conservation
into a spontaneously broken or hidden symmetry.
The rigidity of the phase of the order parameter against local perturbations
and the absence of any phase diffusion gives rise to the Goldstone modes,
which take the form of collision-less (zero)-sound or hydrodynamic sound,
respectively, depending on whether the sound frequency is in the collision-less
mean-field regime or in the collision-dominated regime \cite{0}.

In finite systems, and thus also in all trapped Bose-gases, sharp
phase-transitions are impossible and hidden symmetries in a rigorous sense
cannot
appear. However, a macroscopic wave-function describing a 
Bose-Einstein
condensate still exists
\cite{1}.
Its phase cannot be stable and must undergo a diffusion process, which
restores the $U(1)$ gauge-symmetry over sufficiently long time intervals.
 This
diffusion process is different from the Goldstone modes mentioned
before, which are oscillations around a fixed value of the phase and do
not restore the symmetry. Rather the Goldstone modes show up either as  collision-dominated hydrodynamic phonons or as collision-less phonons, which have
also been observed in the finite Bose-Einstein condensates.
In the present paper I would like to discuss the dynamics of the complex amplitude of a Bose-Einstein condensate containing a finite number of particles, and in particular analyse the diffusion of its phase. My discussion will  extend and correct in several respects the work published in \cite{PRL} .

The stability of the phase-difference between the macroscopic wave functions of two Bose-Einstein condensates in a trap has been measured.
In the experimental set-up  \cite{2}
the relative phase was measured using a time-domain
separated-oscillatory-field condensate interferometer. Over the time-interval
of 100 ms scanned in the experiment the relative phase was found to be robust.
This experimental result demonstrates that the macroscopic wave functions of the condensates cannot be considered as quantum mechanical wave-functions of  many particle systems entangled with each other, whose
decoherence would indeed be extremely rapid. Rather, the macroscopic wave functions are  appropriately viewed as robust classical objects, their quantum mechanical origin (just like magnets, crystals etc) notwithstanding. This does, of course not preclude that there may be quantum effects, for finite condensates, which lead to corrections of the dynamics described by the underlying classical wave-equation, the well-known Gross-Pitaevskii equation \cite{5}.
In a number of papers \cite{3} the dispersion of the phase of
a trapped Bose-Einstein condensate at zero-temperature was considered, which
is due to  fluctuations $\delta\mu$ of the chemical potential
$\mu$ in a finite system with fixed particle number. An extension of this
mechanism to finite temperature has also been proposed \cite{4}. This effect
is not an irreversible phase diffusion but corresponds to an effect of inhomogeneous
broadening, similar to the de-phasing of precessing spins occurring in spin-systems due to inhomogeneous broadening. Like the  decay of the magnetization can be reversed in spin-echos, the decay of the order parameter expectation value in Bose-Einstein condensates due to a finite variance of $\delta\mu$  is in principle reversible in `revivals'. Experiments in Bose-Einstein condensation are
done at  temperature $k_BT\gg\hbar\bar{\omega}$ and often even at
$k_BT\gg\mu$, where
$\bar{\omega}$ is the geometrical mean of the three main trap frequencies.
A process of
phase-diffusion process should occur in such a regime due to the interaction of
the condensate with a thermal bath of collective modes and quasiparticles.
An estimate of this phase-diffusion is of interest for the theory of atom lasers, because the fundamental
limit of the line-width of an atom laser for a given temperature
depends on it similar to the
 `Schawlow-Townes'-formula \cite{4b} for the line-width of a laser.

In this paper  a theory of dissipation and thermal
fluctuations of a trapped Bose-Einstein condensate will be formulated. 
First a phenomenological framework for the theory in
the form of a Langevin equation will be given
in which dissipation appears via a phenomenological parameter and the
fluctuation-dissipation relation is invoked to relate it to maximally three
 intensity-coefficients of the fluctuations. The solution of the Langevin-equation
then determines the relaxation of the condensate-number and the diffusion of the phase, quite similar to the dynamics of a laser-amplitude above threshold.
Then the    Langevin equation is derived from the microscopic theory and formulas for the phenomenological parameters are derived. These are evaluated for a box-like trap and an isotropic harmonic trap-potential as a function of
temperature, particle-number and scattering length.
The final section contains a discussion of our results and a comparison with earlier related work.
The theory presented here makes no claim to apply to the critical regime,
nor can we examine here to what extent it covers the regime below but close to $T_c$, where it may be 
important to take the dynamics of the thermal cloud of non-condensed atoms 
into account in addition to the excitations from the condensate.

\section{Microscopic equations of motion}\label{sec:2}

The weakly interacting Bose-gas in a trap
in standard notation is described by the Hamiltonian
\begin{equation}
 \hat{H}=\int\!d^3\!x
 \hat{\psi}^+\Big\{-\frac{\hbar^2}{2m}\nabla^2+
  V(\bbox{x})-\langle\mu\rangle+\frac{U_0}{2}\hat{\psi}^+
  \hat{\psi}\Big\}\hat{\psi}\,.
 \label{eq:H}
\end{equation}
The total number of atoms $N$ is fixed, i.e. the Hilbert space is the restriction of the Fock-space of $\hat\psi$ to the subspace on which $\hat{N}=N$ is satisfied. $\langle\mu\rangle$ is the average of the  chemical potential, which is a fluctuating quantity in a system where N is fixed. Later-on we shall denote
the fluctuating part of the chemical potential with $\Delta\mu$. The  
presence of a Bose-Einstein condensate in equilibrium means that many ($N_0\gg1$)
particles occupy a single mode of a macroscopic classical matter wave, determined as the mode of lowest energy of the classical Hamiltonian corresponding to eq.(\ref{eq:H}). The latter is obtained by replacing in $H$ the field operator $\hat{\psi}(\bbox x)$ by the classical field $\psi(\bbox x)=\sqrt{\langle N_0\rangle}\exp(i\phi)\tilde\psi_0(\bbox x)$. We shall restrict our attention to sufficiently low temperatures below the critical temperature $T_c$ so that the interaction of the condensate with the mean field of
the thermal cloud of non-condensed particles is negligible. In this way
one finds that the condensate mode $\tilde\psi_0(\bbox{x})$, which we take to be normalized to 1, satisfies the Gross-Pitaevskii equation \cite{5} \begin{equation}
 -(\hbar^2/2m)\nabla^2\tilde\psi_0+
  \big(V(\bbox{x})+U_0\langle N_0\rangle|\tilde\psi_0(\bbox x)|^2\big)\tilde\psi_0=\langle\mu\rangle\tilde\psi_0.
\label{eq:Ha}
\end{equation}
For given $\langle N_0\rangle$ the average value of the chemical potential $\mu$ follows by imposing the normalization condition 
\begin{equation}
\int d^3x|\tilde\psi_0(\bbox x)|^2=1
\end{equation}
on the solution of the Gross-Pitaevskii equation, and thereby $\langle\mu\rangle$ like $\tilde\psi_0(\bbox x)$ 
becomes a function of the mean atom number in the 
condensate $\langle N_0\rangle$. As an important 
consequence of this fact the chemical potential 
of the system can be expressed as a function of 
the average number of atoms in the condensate alone. 
$\langle N_0\rangle$ differs from  $N$, the fixed 
total number of atoms, by the average number 
$\langle N'\rangle$ of non-condensed atoms, 
which needs to be calculated for given
$\langle N_0\rangle$.  The condition 
$N=\langle N_0\rangle+\langle N'\rangle$ then  
fixes $\langle N_0\rangle$ self-consistently. 
In the experimentally realized Bose-Einstein 
condensates it is possible to measure $\langle N_0\rangle$ 
directly with reasonable accuracy as a function 
of temperature, and in practice it is therefore  
reasonable to regard $\langle N_0\rangle$ as an 
experimentally given and known function of temperature. 
The space-dependent mean number density of the 
condensate is $n_0(\bbox{x})=
\langle N_0\rangle|\tilde\psi_0(\bbox{x})|^2$.
We shall  take the mode function $\tilde\psi_0(x)$ in
the Gross-Pitaevskii equation as real and positive. (This also means we are not considering condensates containing vortices). The physical phase of the condensate is not carried by its mode-function $\tilde\psi_0$ but by its complex amplitude denoted as $\alpha_0$, where $\alpha_0=\sqrt{N_0}\exp{i\phi}$.

If $|\alpha_0|^2$ makes a small fluctuation away from
 its equilibrium value $\langle N_0\rangle$
the condensate mode function $\psi_0$ will no longer satisfy eq.(\ref{eq:Ha})
but will change its form slightly. We shall
assume that such fluctuations of $N_0=|\alpha_0|^2$ occur on a sufficiently large time-scale, 
that the new form is again determined
by the Gross-Pitaevskii equation, but for the changed condensate number
$|\alpha_0|^2$ and a correspondingly changed chemical potential $\mu_0$ determined uniquely by $|\alpha_0|^2$, i.e.
in eq.(\ref{eq:Ha}) the replacements $(\tilde\psi_0,\langle N_0\rangle,\langle\mu\rangle)\rightarrow (\psi_0,|\alpha_0|^2,\mu_0)$ have to
be made in this case,
\begin{equation}
 -(\hbar^2/2m)\nabla^2\psi_0+
  \big(V(\bbox{x})+U_0|\alpha_0|^2|\psi_0(\bbox x)|^2\big)\psi_0=\mu_0\psi_0.
\label{eq:Ha1}
\end{equation}
We cannot expect, in general,  that in any given nonequilibrium state the difference defined by $\Delta_0\mu=\mu_0-\langle\mu\rangle$ is the 
{\it total} deviation of
the chemical potential from its equilibrium value, because there may 
obviously be
 states with $|\alpha_0|^2=\langle N_0\rangle$ which differ in
other respects  from the equilibrium state and may therefore have $\mu\ne\langle\mu\rangle$.
Therefore we use the notation $\mu_0$ for the part of the nonequilibrium chemical potential determined by $|\alpha_0|^2$.

The presence of the highly
occupied condensate mode makes  the decomposition of the Heisenberg
field-operator 
\begin{equation}
\hat{\psi}(\bbox{x},t)=\left(|\alpha_0|\exp{(i\phi)}\psi_0(\bbox{x})+
\hat{\chi}(\bbox{x},t)
\right)\exp{(-i\langle\mu\rangle t/\hbar)}
\label{2}
\end{equation}
useful, where we  follow  
 Bogoliubov \cite{7} and describe the condensate
classically. 
$\hat{\chi}(\bbox{x},t)$ is taken to be the
field operator for the particles outside the condensate.   We shall assume  that the temporal changes in $\phi$ can be considered as slow on the time-scales of the dynamics of $\hat{\chi}$. The phase $\phi$ and amplitude $|\alpha_0|$ are additional c-number variables in (\ref{2}). Therefore the taking of expectation values has from now on to include an integration over a distribution of $|\alpha_0|$ and in addition an integration over all values of $\phi$. Since the total number $N$ is fixed $\langle\hat\psi\rangle=0$ must
hold for all times. However, it will also be useful to consider expectation values in the Fock-space of the operators $\hat\chi, \hat\chi^+$ alone without averaging over $\phi$. Such expectation values will be denoted as $\langle ...\rangle_\phi$.

Gauge invariance, strictly speaking, is lost by splitting off a c-number term from the field-operator. However, this symmetry is saved by adopting the rule that the phase $\phi$ of the c-number term in the decomposition also changes under a gauge transformation according to $\phi\rightarrow\phi+\epsilon$. By this device we take into account the fact that the same change of phase would have occurred automatically, if we had not replaced the condensate term by a c-number. The generator of gauge transformations is thus taken as
\begin{equation}
\hat N=i\frac{\partial}{\partial\phi}\Big |_{\hat\chi,\hat\chi^+}
+\int d^3x\hat\chi^+\hat\chi,\label{gauge}
\end{equation}
from which it is clear (cf. eq.(\ref{N})) that $i\frac{\partial}{\partial\phi}|_{\hat\chi,\hat\chi^+}$
is a representation of $\hat{N}_0$.\footnote{This operator with fixed $\hat\chi,\hat\chi^+$ has to be well distinguished from the {\it unrestricted} derivative operator $i(\partial/\partial\phi)$, which is a representation of the {\it total} particle  number $N$ and has as formal
canonical-conjugate $\hat\phi$ with $\exp(i\hat\phi)=\exp(\partial/\partial N)$.} The canonical-conjugate is the phase $\hat\phi$
with 
\begin{equation}\exp(i\hat\phi)=\exp(\partial/\partial N_0)_{\hat\chi,\hat\chi^+}.\label{can}
\end{equation}

Via (\ref{2}) the Hamiltonian furthermore
splits up according to $\hat{H}=H_0+\hat{H_1}+\hat{H_2}+\hat{H_3}+\hat{H_4}$
where  $\hat{H_n}$ 
comprises the terms of $\hat{H}$ which are of n-th order in $\hat\chi,
\hat\chi^+$. 
Explicitly
\begin{equation}
H_0=|\alpha_0|^2\int d^3x\psi_0\{-\frac{\hbar^2}{2m}\nabla^2+V(\bbox x)-\mu_0+\frac{U_0}{2}|\alpha_0|^2|\psi_0|^2\}\psi_0
+(\mu_0-\langle\mu\rangle)|\alpha_0|^2
\label{H_0}
\end{equation}
\begin{equation}
\hat H_1=|\alpha_0|\int d^3x \{ e^{-i\phi}\hat\chi(-\frac{\hbar^2}{2m}\nabla^2+V(\bbox x)-\langle\mu\rangle+U_0|\alpha_0|^2\psi_0^2)\psi_0+(h.c.)\}\label{H_1}
\end{equation}
\begin{eqnarray}
\hat H_2=\int d^3x\{&&\hat\chi^+(-\frac{\hbar^2}{2m}\nabla^2+V(\bbox x)-\mu_0)\hat\chi\nonumber\\
&&+\frac{U_0}{2}|\alpha_0|^2\psi^{2}_0(e^{-2i\phi}\hat\chi^2+e^{2i\phi}\hat\chi^{+2}+4\hat\chi^+\hat\chi)\label{H_2}\\
&&+(\mu_0-\langle\mu\rangle)\hat\chi^+\hat\chi\}\nonumber
\end{eqnarray}
\begin{equation}\hat H_3=U_0|\alpha_0|\int d^3x\psi_0\hat\chi^+(e^{-i\phi}\hat\chi+e^{i\phi}\hat\chi^+)\hat\chi\label{H_3}
\end{equation}
\begin{equation}\hat H_4=\frac{U_0}{2}\int d^3x\hat\chi^+\hat\chi^+\hat\chi\hat\chi.\label{H_4}
\end{equation}   
Using the Gross-Pitaevskii equation (\ref{eq:Ha1}) and its derivative with respect to $|\alpha_0|^2$ we derive in the appendix
\begin{equation}
H_0=\int_{\langle N_0\rangle}^{|\alpha_0|^2}d N_0(\mu_0(N_0)-\langle\mu\rangle)
\label{exp}\end{equation}
The term $\hat{H_1}$ can be simplified using the Gross-Pitaevskii equation (\ref{eq:Ha1}) and becomes then
\begin{equation}
\hat H_1=|\alpha_0|(\mu_0-\langle\mu\rangle)\int d^3x ( e^{-i\phi}\hat\chi+e^{i\phi}\hat\chi^+)\psi_0\label{H_1'}
\end{equation}
This expression will be seen to vanish below due to an orthogonality condition.

The first part of $\hat{H_2}$ in (\ref{H_2}) is  diagonalized by introducing
quasi-particle operators $\hat{\alpha}_{\nu},\hat{\alpha}_{\nu}^{+}$ defined by 
the standard Bogoliubov transformation, with time-dependent $\phi(t)$ 
\begin{equation}
\hat\chi(\bbox{x})=e^{i\phi}\sum_\nu
\big(u_\nu(\bbox{x})\hat{\alpha}_\nu+
v^*_\nu(\bbox{x})\hat{\alpha}_\nu^+\big)\,.\label{Bog}
\end{equation}
The $u_\nu$, $v_\nu$ satisfy the usual Bogoliubov-Fetter equations
\begin{equation}
 \left(\begin{array}{cc}
         -\frac{\hbar^2}{2m}\nabla^2+U_{\rm eff}(\bbox{x})-\hbar\omega_\nu &
         K(\bbox{x})\\
          K(\bbox{x}) &
            -\frac{\hbar^2}{2m}\nabla^2+U_{\rm eff}(\bbox{x})+\hbar\omega_\nu
             \end{array}\right)
             {u_\nu(\bbox{x}) \choose v_\nu(\bbox{x})}=0,
\label{eq:2.18}
\end{equation}
with the abbreviations
\begin{eqnarray}
 &&  U_{\rm eff}(\bbox{x})=V(\bbox{x})-\mu_0
  +2U_0 |\alpha_0|^2 \psi_0(\bbox{x})^2\nonumber\\
 && K(\bbox{x})=|\alpha_0|^2 U_0\psi_0(\bbox{x})^2\,.
 \label{eq:2.19}
\end{eqnarray}
The Hamiltonian $\hat H_2$ now takes the form
\begin{equation}
\hat H_2=\sum_\nu\hbar\omega_\nu(\hat\alpha_\nu^+\hat\alpha_\nu+|v_\nu|^2)+(\mu_0-\langle\mu\rangle+\hbar\dot\phi)\int d^3x\hat\chi^+\hat\chi
\label{H_2'}
\end{equation}
The coefficients $u_\nu, v_\nu$ and the mode frequencies $\omega_\nu$
become also functions of $|\alpha_0|$ and fluctuate (slowly) with that number. Their equilibrium values will be denoted by $\tilde u_\nu, \tilde v_\nu$
and $\tilde\omega_\nu$ and the corresponding operator $\hat\chi$
according to (\ref{Bog}) as $\hat{\tilde\chi}$. 

Eq.~(\ref{eq:2.18}) is consistent with  the ortho-normality conditions
\begin{eqnarray}
 && \int d^3x(u_\nu u_\mu^*-v_\nu v_\mu^*)=\delta_{\nu \mu}\label{12a}\\
 && \int d^3r(u_\nu^*v_\mu-u_\mu^*v_\nu)=0
\label{eq:2.20}
\end{eqnarray}
which guarantee the Bose commutation relations of the $\alpha_\nu$,
$\alpha_\mu^+$.
A formal solution of eq.~(\ref{eq:2.18}) at zero energy $\hbar\omega_\nu=0$ is given
by the condensate
\begin{equation}
 u_\nu(\bbox{x})=-v_\nu^*(\bbox{x})=\psi_0(\bbox{x})\quad,\quad \omega_\nu=0\,,
\label{eq:2.21}
\end{equation}
This solution is obviously not normalizable in the required sense (\ref{12a})
to furnish
an acceptable solution for the $u_\nu , v_\nu$ and must therefore be excluded from the sum over the terms containing the operators $\hat{\alpha}_\nu, \hat{\alpha}^+_\nu$. The existence of this formal solution implies however 
that the properly normalizable solutions $u_\nu, v_\nu$ and the condensate mode
$\psi_0$ satisfy the important orthogonality relation \begin{equation}\int d^3x\psi_0(u_\nu+v_\nu)=0.\end{equation} 
It follows with (\ref{Bog}) that
\begin{equation}
\int d^3x\psi_0(e^{-i\phi}\hat\chi+e^{i\phi}\hat\chi^+)=0\label{p}
\end{equation}
which in turn implies that the reduced expression (\ref{H_1'})
for $\hat H_1$ vanishes.
Using the property (\ref{p}) one can verify that the decomposition (\ref{2}) of $\hat{\psi}$   implies  
\begin{equation}
N=|\alpha_0|^2+\hat{N'}\label{N_0}\label{N}
\end{equation}
with
\begin{eqnarray}
\hat{N'}=\int d^3x\hat\chi^+&&(\bbox x)\hat\chi(\bbox x)\nonumber\\
=\sum_{\nu,\mu}\int d^3x&&\big(\hat{\alpha}_\nu^+\hat{\alpha}_\mu(u_\nu^* u_\mu+v_\nu^* v_\mu)+\frac{1}{2}\hat{\alpha}_\nu\hat{\alpha}_\mu (u_\nu v_\mu+v_\nu u_\mu)\nonumber\\
+&&\frac{1}{2}\hat{\alpha}^+_\nu\hat{\alpha}^+_\mu (u^*_\nu v^*_\mu+v^*_\nu u^*_\mu)+\delta_{\nu\mu}|v_\nu|^2\big)
\end{eqnarray}
which serves as a definition of $N_0=|\alpha_0|^2$. 
The mean thermal density $n'$ in equilibrium can now be determined
via
\begin{equation}
n'(\bbox x)=\langle\hat{\tilde\chi}^+(\bbox x)\hat{\tilde\chi}(\bbox x)\rangle =\sum_\nu\big\{(|\tilde u_\nu(\bbox x)|^2+|\tilde v_\nu(\bbox x)|^2)\bar{n}_\nu+|\tilde v_\nu(\bbox x)|^2\big\}
\end{equation}
with $\bar n_\nu=(\exp(\beta\hbar\tilde\omega_\nu)-1)^{-1}$.

The fluctuations of $N_0$ are similarly fixed by
\begin{eqnarray}
\langle\Delta N_0^2\rangle  =
\langle N_0^2\rangle-\langle &&N_0\rangle^2 = \langle\Delta \hat{N}'^2\rangle \label{deln}\\
 = \sum_\nu\sum_{\nu'}\Bigg \{&&\bar{n}_\nu(\bar{n}_{\nu'}+1)\left|\int d^3x(\tilde u^*_\nu(x)\tilde u_{\nu'}(x)+\tilde v_{\nu'}(x)\tilde v^*_\nu(x))\right|^2\nonumber\\
&& +(\bar{n}_\nu \bar{n}_{\nu'}+\frac{1}{2}(\bar{n}_\nu+\bar{n}_{\nu'}+1))\left|\int d^3x(\tilde u_\nu(x)\tilde v_{\nu'}(x)+\tilde u_{\nu'}(x)\tilde v_\nu(x))\right|^2\Bigg\}
\label{deltan}
\end{eqnarray}
They have been evaluated in \cite{11a} and are also needed below (see
eq.(\ref{Delta})). For work in the mathematical physics literature on
number-fluctuations in the condensate of the ideal Bose-gas and models of the interacting Bose-gas see \cite{V1,V2} and references given there.
For an alternative proposal to define and calculate the number-fluctuations in a Bose-condensate see \cite{W}.

After the transformation (\ref{Bog}) the Hamiltonian is now in the form
\begin{equation}
\hat H= H_0+\hat H_2+\hat H_3+\hat H_4\label{neu}
\end{equation}
with $H_0, \hat H_2, \hat H_3, \hat H_4$ given by eqs.(\ref{exp},\ref{H_2'},\ref{H_3},\ref{H_4}). 

\section{Langevin-equation of the condensate amplitude}\label{sec:3}

Neither the Gross-Pitaevskii equation nor the Bogoliubov-Fetter equations furnish an equation for the condensate amplitude $\alpha_0=\sqrt{N_0}\exp{i\phi}$.
To find such an equation phenomenologically we first
turn to a macroscopic quantity like the  entropy $S(|\alpha_0|^2,N)$ for a fixed particle number N, but restricted to a fixed arbitrary value of $\alpha_0=\sqrt{N_0}\exp(i\phi)$, where $N_0$ is the instantaneous number of particles in the condensate and different from the equilibrium value $\langle N_0\rangle$ corresponding to the maximum of $S(|\alpha_0|^2,N)$. Thus $\langle N_0\rangle$ is a function of N.
 The fluctuations of $N_0$ in the closed system formed by the trapped condensate after the evaporative cooling has been switched off are governed by a canonical Boltzmann-Einstein distribution 
\[
P(N_0)=\Omega^{-1}\exp{\big (S(|\alpha_0|^2,N)/k_B\big )}.
\] 
We shall restrict ourselves to temperatures in the condensed regime outside the critical regime, where $\langle N_0(N)\rangle $ is much larger than its root mean square $\sqrt{\langle\Delta N_0^2(N)\rangle} =\sqrt{\langle\Delta \hat{N'}^2\rangle} =(\langle\hat{N'}^{2}\rangle-\langle\hat{N'}\rangle^2)^{1/2}$, which is also a function of N. Then $S(|\alpha_0|^2,N)$, expanded to  lowest order around its maximum,  takes the form

\[  
S(|\alpha_0|^2,N)=S^{(eq)}(N) + \Delta S(|\alpha_0|^2,N)
\]
with
\begin{equation}
\Delta S(|\alpha_0|^2,N)=-k_B\frac{(|\alpha_0|^2-\langle N_0\rangle)^2}{2\langle\Delta  
N_0^2\rangle}\label{DeltaS}
\end{equation}
The entropy $S(|\alpha_0|^2,N)$ not only determines the equilibrium-distribution of the condensate amplitude, but
appears also in its equation of motion, both in the conservative part of the dynamics as a conserved quantity, and in the dissipative part
as a potential for the irreversible part of the dynamics. 
Let us first consider both parts separately. 

The conservative part of the dynamics of $\alpha_0$ is connected with the dynamics of its phase $\phi$. According to eqs.(\ref{2},\ref{Bog}) a change of $\phi$ changes the total phase of the 
field-operator $\hat\psi$. For this reason the dynamics of $\phi$ is given
by the equation of motion
\begin{equation} \dot{\phi}=-\frac{1}{\hbar}\frac{\partial \langle\hat H\rangle}{\partial N}=-\frac{1}{\hbar}\Delta\mu
\label{X}
\end{equation}
where $\Delta\mu$ is the deviation of the chemical potential from its equilibrium value. Such deviations may occur as a result of any fluctuations
present in the system and, as discussed already, 
may in particular occur as a result of fluctuations 
of the value of $N_0$ away from its average $\langle N_0\rangle$. 
This part of the fluctuation of $\mu$ we shall denote as $\Delta_0\mu$.
 Expanding again to lowest order around the equilibrium $N_0=\langle N_0\rangle$ we can write 
\begin{equation}
\Delta_0\mu=\frac{\partial\langle\mu\rangle}{\partial \langle N_0\rangle}(|\alpha_0|^2-\langle N_0\rangle)
\label{28a}
\end{equation}
The systematic part of the 
conservative part of the equation of motion of $\alpha_0$ 
can now be written in the form
\begin{equation} \big(i\hbar\dot{\alpha}_0\big)_{cons}=\Delta_0\mu \alpha_0.\label{cons}
\end{equation}

It is convenient to introduce the fluctuation of the free energy by
 $\Delta F= -T\Delta S$. The dynamics (\ref{cons}) conserves $|\alpha_0|^2$ and $\Delta F$. In equilibrium the right-hand side of this equation vanishes, 
because there $\Delta_0\mu=0$, 
and the total phase of the condensate 
$\phi-\langle\mu\rangle t/\hbar$ changes only 
with a rate given by the {\it average} chemical potential 
$\langle\mu\rangle$ in equilibrium.

The dissipative part of the equation of motion of $\alpha_0$ near thermal  
equilibrium is written with the help of $\Delta F$ in the  
general form 
\begin{equation}
 \hbar\big (\dot{\alpha}_0\big )_{diss}=-\Gamma_0
  \frac{\partial \Delta F(|\alpha_0|^2,N)}{\partial\alpha_0^*}
\label{eq:diss}
\end{equation}
which contains the positive phenomenological parameter $\Gamma_0$ and 
describes the relaxation of $N_0=|\alpha_0|^2$ to its equilibrium-value
$\langle N_0\rangle$.

According to general principles of statistical thermodynamics \cite{8}
the relaxation process (\ref{eq:diss}) must be accompanied by some form of noise. Adding a noise-term the total Langevin-equation of $\alpha_0$ can be written in the form
\begin{equation}
 i\hbar\dot{\alpha}_0=\Delta_0\mu \alpha_0-i\Gamma_0
  \frac{\partial \Delta F(|\alpha_0|^2,N)}{\partial\alpha_0^*}+\xi(t)\exp(i\phi).
\label{eq:1}
\end{equation}
Since the condensate amplitude $\alpha_0$ is a collective quantity the noise $\xi(t)$ can be assumed to be Gaussian due to the central limit theorem. In addition we shall assume $\xi(t)$ to be a white noise force. This means that the actual correlation time $\tau_m$ of the noise $\xi$ is assumed
to  be much smaller than the time-scale on which the dynamics of $\alpha_0$ is observed, an assumption which must be checked for its validity in any concrete microscopic description. (In the microscopic theory we describe later it
is a consistent assumption because the relaxation rate $\gamma_c$ of $|\alpha_0|^2$ turns out to be small compared to the time-scale of motion
in the trap). Thus we assume that  $
 \langle \xi(t)\rangle=0$ and
\begin{eqnarray}
\langle \xi^*(t)\xi(0)\rangle=
  &&\hbar k_BT(2\Gamma_0+\Gamma')\delta(t)\label{C1}\\
\langle \xi(t)\xi(0)\rangle=
  &&\hbar k_BT(\Gamma'+i\Gamma'')\delta(t)\label{C2}
\end{eqnarray}
where $\Gamma_0$ reappears in (\ref{C1}) because of the
fluctuation-dissipation theorem. The form of the Langevin-equation (\ref{eq:1}) generalizes the work in \cite{PRL}  by taking into account a possible correlation of the phase of the condensate and of the Langevin force, which may exist in condensates with finite particle numbers due to gauge invariance, i.e. particle-number conservation. (However, it will turn out later that the coefficient $\Gamma''$ vanishes in condensates with a real condensate mode
i.e. without vortices,
which can be understood generally as a consequence of time-reversal symmetry.)
Gauge-invariance implies that the Langevin-equation for $\alpha_0$, including the fluctuating term, must be invariant under the transformation $\phi\rightarrow\phi+\epsilon$. This makes it useful to write  the fluctuating term as $\exp(i\phi(t))\xi(t)$ where $\xi(t)$ is a complex noise source
{\it independent} of $\phi$, which, physically, describes the scattering of particles in the condensate with those outside.\footnote{A simpler ansatz (see\cite{PRL}) ignores the $\phi$-dependence of the Langevin-force in (\ref{eq:1}). Then gauge-invariance of the Fokker-Planck equation which is
stochastically equivalent  to the Langevin equation implies $\Gamma'=\Gamma''=0$, i.e. the complex noise $F_0(t)$ then has random phase-fluctuations which are completely uncorrelated with and equi-distributed with respect to the condensate-phase $\phi$. Note however that this achieves gauge-invariance only in an ensemble sense, not for each individual stochastic physical realization which together form the ensemble. In contrast
 the form of the Langevin equation
considered here does enforce
 gauge invariance for each stochastic realization.} 
The  coefficients $\Gamma', \Gamma''$  describe a possible correlation of the phases of $F_0$ and $\alpha_0$, i.e. the existence of a squeezing in the thermal bath of uncondensed 
particles, caused by the constraint of total particle-number conservation.
 We shall see that this  effect actually does occur in finite condensates, i.e. the condensate mode imprints its (slowly) fluctuating phase on the non-condensed 'environment' due to particle number conservation
in such a way that the lowest-lying modes are nearly {\it totally} squeezed.

The multiplicative nature of the noise in (\ref{eq:1}) raises the question in which stochastic calculus this equation should be interpreted: in the sense of Ito, or Stratonovich, or in some intermediate sense? This will be specified in a moment. Within the Gauss- and Markoff-assumption the form of the noise-force  with 
the same positive coefficient $\Gamma_0\geq 0$ appearing in the dissipative 
part (\ref{eq:diss}) and two further real coefficients $\Gamma', \Gamma''$,  is fixed by the requirement that the Langevin-equation must be consistent with the correct equilibrium distribution \cite{8}
$P(\alpha_0,\alpha_0^*)=Z^{-1}\exp(-\Delta F(|\alpha_0|^2,N)/k_BT)$
for the condensate. 
Splitting into real and imaginary parts eqs.(\ref{C1},\ref{C2}) become
\begin{eqnarray}
\langle\Re(\xi(t))\Re(\xi(0))\rangle=&&\hbar k_BT(\Gamma_0+\Gamma' )\delta(t)
\label{R1}\\
\langle\Im(\xi(t))\Im(\xi(0))\rangle=&&\hbar k_BT\Gamma_0\delta(t)
\label{R2}\\
\langle\Re(\xi(t))\Im(\xi(0))\rangle=&&\frac{1}{2}\hbar k_BT\Gamma'' \delta(t).
\label{R3}
\end{eqnarray}
Eq.(\ref{eq:1}) may now be rewritten as
\begin{eqnarray}
\frac{\partial N_0}{\partial t}=&&-2\frac{\Gamma_0}{\hbar}(N_0\frac{\partial\Delta F}{\partial N_0}-k_BT)+\frac{2}{\hbar}\sqrt{N_0}\Im(\xi(t))\label{26a}\\
\frac{\partial\phi}{\partial t}=&&-\frac{1}{\hbar}\Delta_0\mu-\frac{1}{\hbar\sqrt{N_0}}\Re(\xi(t))
\label{26b}\end{eqnarray}
and must in this form be interpreted as a stochastic differential equation
in the sense of Ito\footnote{Then the Fokker-Planck equation corresponding to eqs.(\ref{26a},\ref{26b}) is 
$\hbar\partial P/\partial t = 2\Gamma_0\partial/\partial N_0 [N_0(\partial\Delta F/\partial N_0  +k_BT\partial/\partial N_0 )P]$ and has the desired equilibrium distribution $\sim \exp(-\Delta F/k_BT)$. 
This implies that eq.(\ref{eq:1}) must be interpreted in some intermediate sense, which we need not specify here
further. In order to obtain its  version in the sense of Ito it is best to bring (\ref{26a}),(\ref{26b}) into the form (\ref{eq:1}) using the Ito calculus.}. 

Eq.(\ref{26b}) can be compared  with (\ref{X}). This comparison reveals that $\Re(\xi(t))$ must describe  the fluctuations of the chemical potential {\it
not} caused by
deviations of $|\alpha_0|^2$ from its equilibrium value but by other fluctuations in the system. We shall come back to this point in section \ref{sec.5a} below. 

The 
three phenomenological coefficients $\Gamma_0, \Gamma', \Gamma''$, 
are dimensionless, temperature dependent numbers, which must be determined from a microscopic theory. Only one of these coefficients, $\Gamma_0$, is connected with the fluctuations of the number of condensed atoms.
If fluctuations of the chemical potential due to other processes
are neglected, i.e. $\Re{\xi(t)}=0$,
 the remaining two coefficients are fixed at
\begin{equation}
\Gamma'=-\Gamma_0, \qquad \Gamma''=0
\end{equation}
This corresponds to the case of maximal squeezing of the noise in the
phase-direction.

Linearizing with respect to the small fluctuations $\delta N_0\ll\langle N_0\rangle$ we find
\begin{equation}\hbar\delta \dot{N}_0=-\frac{2k_BT}{\langle\Delta N_0^2\rangle}\langle N_0\rangle\Gamma_0\delta N_0+2\sqrt{\langle N_0\rangle}\Im(\xi(t))\label{30}
\end{equation}
\begin{equation}
\hbar \dot{\phi}=-\frac{\partial \langle \mu\rangle}{\partial \langle N_0\rangle}\delta N_0-\frac{1}{\sqrt{\langle N_0\rangle}}\Re(\xi(t))\label{30a}
\end{equation}
Eq.(\ref{30}) describes
the relaxation of the condensate to the equilibrium at $\langle N_0\rangle=\langle|\alpha_0|^2\rangle$ on the time-scale 
\begin{equation}
\tau_c=\frac{\hbar \langle\Delta N_0^2\rangle}{2\Gamma_0\langle N_0\rangle k_BT}\label{tau}
\end{equation}
 and the thermal fluctuations around it with the correlation function
\begin{equation}
 \langle\delta N_0(t)\delta N_0(t')\rangle = \langle\Delta N_0^2\rangle
  e^{-|t-t'|/\tau_c}
\label{eq:3a}
\end{equation}
The correlation time
$\tau_c$  
is an important time-scale of the problem.
The noise sources $\Im{(\xi(t))}, \Re{(\xi(t))}$ must have correlation times short compared to $\tau_c$ in order to be well described by white noise.

 On a time-scale  much larger
than the correlation time $\tau_c$ the fluctuations $\delta N_0(t)$ 
in the equation for the phase can also be  considered as Gaussian white noise with correlation function $2\tau_c\langle\Delta N_0^2\rangle\delta(t)$. Using this long-time approximation in the equation for the phase $\phi$ and taking the correlation of the effective white noise $\delta N_0(t)$ with $\Re(\xi(t))$ properly into account, $\phi(t)$ is found to satisfy the Langevin equation of a Wiener process
\[
d\phi(t)=\sqrt{D_\phi}dw
\]
with $(dw)^2=dt$ and diffusion constant
\begin{equation}
 D_\phi= \frac{\langle\Delta N_0^2\rangle}{\hbar\langle N_0\rangle}\frac{\partial\langle\mu\rangle}{\partial\langle N_0\rangle}\left(\frac{\langle\Delta N_0^2\rangle}{k_BT}\frac{\partial \langle\mu\rangle}{\partial \langle N_0\rangle}\frac{1}{\Gamma_0}+
\frac{\Gamma''}{\Gamma_0}\right)+\frac{k_BT}
{\hbar\langle N_0\rangle}(\Gamma_0+\Gamma'),
\label{eq:PD}
\end{equation}
i.e 
\begin{equation}\langle\big(\phi(t)-\phi(0)\big)^2\rangle=D_\phi |t|.\label{34a}
\end{equation}
Eq.(\ref{eq:PD}) agrees with the result of \cite{PRL} if we assume
as in \cite{PRL} no squeezing in the noise, i.e. $\Gamma'=\Gamma''=0$ and
$\partial\langle\mu\rangle/\partial\langle N_0\rangle=k_BT/\langle\Delta N_0^2\rangle^{-1}$. Both assumptions will not be made in the present work, however (cf. also the corresponding discussion in the final section).

The expectation value $\langle\alpha_0(t)\rangle$ then decays exponentially
according to $\langle\alpha_0(t)\rangle=\sqrt{\langle  
N_0\rangle}e^{-\Delta\nu |t|}$ with the
line-width $\Delta\nu$ given by the Schawlow-Townes-type formula
$ \Delta\nu=\frac{1}{2}D_\phi.$

It is not difficult to solve eqs.(\ref{30},\ref{30a}) for the phase-fluctuations also on
time-scales of the order of $\tau_c$. The result for the mean square of the phase-increment in time is
\begin{equation}
\langle(\phi(t)-\phi(0))^2\rangle=D_\phi|t|+2\frac{\partial\langle\mu\rangle}{\partial\langle N_0\rangle}\left(k_BT\Gamma''+\langle\Delta N_0^2\rangle
\frac{\partial\langle\mu\rangle}{\partial\langle N_0\rangle}\right)\tau_c^2\big(e^{-|t|/\tau_c}-1\big).
\end{equation}
It interpolates between the diffusive long-time behavior (\ref{34a}) for $t>>\tau_c$ and the short-time behavior for $t\ll\tau_c$
\begin{equation}
\langle(\phi(t)-\phi(0))^2\rangle=\frac{k_BT}{\hbar\langle N_0\rangle}(\Gamma_0+\Gamma')|t|+\frac{1}{\hbar^2}
\frac{\partial \langle\mu\rangle}{\partial\langle N_0\rangle}\left(\frac{\partial \langle\mu\rangle}{\partial\langle N_0\rangle}\langle\Delta N_0^2\rangle+k_BT\Gamma''\right)t^2.
\end{equation}
The first term describes phase diffusion due to thermal fluctuations of the chemical potential on time scales much shorter than $\tau_c$. The second term describes a non-diffusive and in principle reversible phase-collapse \cite{3,4} with the collapse rate
\begin{equation}
\gamma_{collapse}=\frac{1}{\hbar}\sqrt{\frac{\partial\langle\mu\rangle}{\partial\langle N_0\rangle}\left(\frac{\partial\langle\mu\rangle}{\partial\langle N_0\rangle}\langle\Delta N_0^2\rangle+k_BT\Gamma''\right)}\label{coll} 
\end{equation}
including a contribution from the cross-correlation 
between both types of fluctuations.
 
\section{Microscopic derivation of the Langevin equation}\label{sec:4}

The microscopic derivation of the equation of motion for the condensate amplitude $\alpha_0$ 
can be carried out by using the Hamiltonian (\ref{neu}). As
we did for the phenomenological equations in the preceding section
we wish to derive here the microscopic equation of motion only to first order in the deviation $(|\alpha_0|-\sqrt{\langle N_0\rangle})$ from equilibrium.

$H_0$ given by eq.(\ref{exp}) is the Hamiltonian, in mean field approximation, of the pure condensate. Its free equation of motion is 
\begin{equation}
i\hbar\dot\alpha_0=\frac{\partial H_0}{\partial\alpha_0^*}=(\mu_0-\langle\mu\rangle)\alpha_0,
\end{equation}
from which
\begin{equation}
\frac{d\phi(t)}{dt}=-(\mu_0(|\alpha_0(t)|^2)
-\langle\mu\rangle)/\hbar\label{59d}
\end{equation}
follows. Let us first use (\ref{59d}) in (\ref{H_2'}) to 
simplify $\hat H_2$ and then eliminate  $\hat H_2$ by proceeding to the Heisenberg picture with respect to it.
This changes $\hat\chi, \hat\chi^+$ in 
$\hat H_3, \hat H_4$ according to 
\begin{equation}
\hat\chi\rightarrow
\hat\chi(t)=e^{i\phi(t)}\sum_\nu(u_\nu\hat\alpha_\nu e^{-i\omega_\nu t}+v^*_\nu\hat\alpha_\nu^+e^{i\omega_\nu t})\label{Bog'}
\end{equation}
 and its adjoint\footnote{For simplicity we disregard here the slow time-dependence of the
frequencies $\omega_\nu$.}.
The transformed time-dependent Hamiltonians  will be denoted as 
$\hat H_3(t), \hat H_4(t)$, but $\hat H_4(t)$ will not be needed in the following.

The equation of motion of the condensate amplitude $\alpha_0$ now takes the form,\footnote{For the canonically conjugate pair $N_0, \phi$ at fixed $\hat\chi,\hat\chi^+$ cf. eqs.(\ref{gauge},\ref{can})} with the notation $\hat H(t)= H_0+\hat H_3(t)+\hat H_4(t)$,
\begin{equation}
 \hbar\frac{d\phi}{dt}=-\left(\frac{\partial \hat H(t)}{\partial|\alpha_0|^2}\right)_{\phi,\hat\chi,\hat\chi^+}\label{mic1}
\end{equation}
\begin{equation}
 \hbar\frac{d|\alpha_0|^2}{dt}=\left(\frac{\partial \hat H(t)}{\partial\phi}\right)_{|\alpha_0|,\hat\chi,\hat\chi^+}\label{mic2}
\end{equation}
We obtain 
\begin{eqnarray}
\hbar \frac{d\phi}{dt}&&=-\Delta_0\mu -\frac{1}{\sqrt{\langle N_0\rangle}}\Re{(\hat\xi'(t))}-\frac{1}{\sqrt{\langle N_0\rangle}}\delta\Re{(\hat\xi'(t))} \label{58a}\\
\hbar\frac{d |\alpha_0|^2}{dt}&&=2\sqrt{\langle N_0\rangle}\Im{(\hat\xi(t))}+2\sqrt{\langle N_0\rangle}\delta\Im{(\hat\xi(t))}\label{58b}
\end{eqnarray}
with
\begin{eqnarray}
\Re({\hat\xi'(t)})=&&\frac{1}{2}U_0
\int d^3x(\tilde\psi_0+2\langle N_0\rangle\frac{\partial\tilde\psi_0}{\partial\langle N_0\rangle})\hat{\tilde\chi}^+(t)(e^{-i\phi}\hat{\tilde\chi}(t)
+e^{i\phi}\hat{\tilde\chi}^+(t))
\hat{\tilde\chi}(t)\label{59a}\\
\Im({\hat\xi(t)})=&&\frac{1}{2i}U_0\int d^3x\tilde\psi_0\hat{\tilde\chi}^+(t)(e^{-i\phi}\hat{\tilde\chi}(t)
-e^{i\phi}\hat{\tilde\chi}^+(t))
\hat{\tilde\chi}(t)\label{59b}
\end{eqnarray}
The complex noise $\xi(t)$ in eqs.(\ref{eq:1}-\ref{26b}) should be identified with
\begin{equation}
\hat\xi(t)=\Re(\hat\xi'(t))+i\Im(\hat\xi(t)).\label{xi'}
\end{equation}
It is indeed independent of $\phi$ as required by gauge-invariance of the Langevin-equation, as can be seen from (\ref{58a}), (\ref{58b}) with (\ref{Bog'}).We shall see in the next section that $\hat\xi'(t)$ can be replaced by a c-number.

To describe fluctuations around equilibrium we have replaced in the preceding expressions the quantities $|\alpha_0|^2, \psi_0, \hat\chi, \hat\chi^+$ by their equilibrium expressions $\sqrt{\langle N_0\rangle}, \tilde\psi_0, \hat{\tilde\chi}, \hat{\tilde\chi}^+$ and represent the difference in the nonequilibrium state by $\delta\Re{(\hat\xi'(t))}, \delta\Im{(\hat\xi(t))}$ 
in eqs.(\ref{58a},\ref{58b}). Omitting these differences altogether
amounts to
neglecting the back-action of the condensate on the thermal reservoir,
which describes not only a modification of the fluctuating forces,
which can indeed be neglected for fluctuations around a
stable thermodynamic equilibrium, but also dissipation.
To take the latter into account we need to calculate the averages
$\delta\langle\delta\Re{(\hat\xi'(t))}\rangle_\phi, \delta\langle\delta\Im{(\hat\xi(t))}\rangle_\phi$ to lowest order in the interaction
between the condensate and the thermal cloud of atoms. The form which these quantities must take is prescribed completely by the fluctuation-dissipation theorem and symmetry:

For the reversible phase-dynamics the back-action can only lead to a shift in the 
average chemical potential. Such shifts due to the interaction $\hat H_3$
will be small and are neglected here. For the irreversible
amplitude dynamics the fluctuation-dissipation theorem  requires in addition the appearance
of a dissipation term. If
\begin{equation}
S_{JJ}(t-t^{\prime})=\langle\Im(\hat\xi(t))\Im(\hat\xi(t')
\rangle_\phi\label{59c}
\end{equation}
is the correlation function of the fluctuating force in(\ref{58b}), the back action must modify eq.(\ref{58b}) to the form 
\begin{eqnarray}
       \hbar\frac{d|\alpha_0|^2}{dt} =
       -\frac{4\langle N_0\rangle}{\hbar k_BT}\int^t_{-\infty}dt^{\prime}
       S_{JJ}(t-t^{\prime})\frac{\partial H_0 (t^{\prime})}
       {\partial|\alpha_0|^2} + 2\sqrt{\langle N_0\rangle}\,\,
       \Im(\hat\xi(t))\label{eq:(F)}
 \end{eqnarray}
The derivation of this equation is given in appendix \ref{sec:11}.

This stochastic differential equation still differs from the phenomenological equation (\ref{30}) in two respects:
\begin{enumerate}
 \item[i)]
      The noise still has a finite correlation time $\tau_{mic}$. We shall
      consider these correlation functions in more detail below. Taking the
      Markovian limit $\tau_{mic}\rightarrow 0$ with

\begin{eqnarray}
      S_{JJ}(t-t^{\prime}) = \hbar k_B T\Gamma_0 \delta(t-t^{\prime})
\end{eqnarray}
eq.(\ref{eq:(F)}) becomes
\begin{eqnarray}
       \frac{d|\alpha_0|^2}{dt} =
       - 2\frac{\Gamma_0}{\hbar}\langle N_0\rangle \frac{\partial H_0}
       {\partial|\alpha_0|^2} +
       \frac{2}{\hbar} \sqrt{\langle N_0\rangle}\,\,\Im(\hat\xi(t))\label{eq:(G)}
\end{eqnarray}
 \item[ii)]
      The mean-field Hamiltonian $H_0(|\alpha_0|^2)$ appears in 
      eqs.(\ref{eq:(F)}),(\ref{eq:(G)}) instead of the free energy $\Delta F (|\alpha_0|^2)$. This is due to the fact that the influence of the thermal excitations on the energy are not yet taken into account. Doing this under isothermal or closed-system boundary conditions we should replace 
        the energy
       $H_0(|\alpha_0|^2)$ by the free energy
$\Delta F (|\alpha_0|^2)$ or $-T\Delta S(|\alpha_0|^2)$, respectively. 
\end{enumerate}
This completes our derivation of the Langevin equation for the complex
 amplitude of the condensate.

\section{Green-Kubo expressions for the transport coefficients}\label{sec:5}

Let us now analyse the fluctuating forces in more detail.
Inserting the Bogoliubov transformation (\ref{Bog'}) in (\ref{59a}, \ref{59b})
  the fluctuating forces  
 take the form
\begin{eqnarray}
\Re({\hat\xi'(t)})= 
 &&\frac{1}{4}\sum_{\kappa\nu\mu}\left(
\left((M'^{(1)}_{\kappa,\nu\mu})^*+M'^{(2)}_{\nu\mu,\kappa}\right)
\hat{\alpha}_\nu^+\hat{\alpha}_\mu^+\hat{\alpha}_\kappa
e^{-i(\tilde\omega_\kappa-\tilde\omega_\nu-\tilde\omega_\mu)t}+
h.c.\right)\nonumber\\
&&+(nonresonant \quad terms).\label{Re}\\
 \nonumber\\
\Im({\hat\xi(t)})= 
 &&\frac{1}{4i}\sum_{\kappa\nu\mu}\left(
\left((M_{\kappa,\nu\mu}^{(1)})^*-M_{\nu\mu,\kappa}^{(2)}\right)
\hat{\alpha}_\nu^+\hat{\alpha}_\mu^+\hat{\alpha}_\kappa
e^{-i(\tilde\omega_\kappa-\tilde\omega_\nu-\tilde\omega_\mu)t}-
h.c.\right)\nonumber\\
&&
+(nonresonant \quad terms)\label{Im}.
 \end{eqnarray}
Terms  are called 'nonresonant' if the frequencies of the quasi-particles cannot add up to zero. Such terms have not been written out explicitly because
later-on we shall restrict ourselves to the resonance or rotating wave approximation in
which they don't contribute. The relevant matrix elements $M^{(1)},M^{(2)}$ are
\begin{eqnarray}
M^{(1)}_{\kappa,\nu\mu}=&&2U_0\int d^3x\tilde\psi_0\tilde v_\nu(\tilde u^*_\kappa 
\tilde u_\mu+\frac{1}{2}\tilde v^*_\kappa \tilde v_\mu)+(\nu \leftrightarrow\mu)\nonumber\\
 \label{matrixel}\\
M^{(2)}_{\nu\mu,\kappa}=&&2U_0\int d^3x\tilde\psi_0\tilde u^*_\nu (\tilde v^*_\mu 
\tilde v_\kappa +\frac{1}{2}\tilde u^*_\mu \tilde u_\kappa)+(\nu \leftrightarrow\mu)\nonumber
 \end{eqnarray}
and very similarly
\begin{eqnarray}
M'^{(1)}_{\kappa,\nu\mu}=&&2U_0\int d^3x(\tilde\psi_0+2\langle N_0\rangle\frac{\partial\tilde\psi_0}{\partial\langle N_0\rangle)})\tilde v_\nu(\tilde u^*_\kappa 
\tilde u_\mu+\frac{1}{2}\tilde v^*_\kappa \tilde v_\mu)+(\nu \leftrightarrow\mu)\nonumber\\
 \label{tildematrixel}\\
M'^{(2)}_{\nu\mu,\kappa}=&&2U_0\int d^3x(\tilde\psi_0+2\langle N_0\rangle\frac{\partial\tilde\psi_0}{\partial\langle N_0\rangle)})\tilde u^*_\nu (\tilde v^*_\mu 
\tilde v_\kappa +\frac{1}{2}\tilde u^*_\mu \tilde u_\kappa)+(\nu \leftrightarrow\mu)\nonumber
 \end{eqnarray}
The matrix-elements $M'^{(1)}, M'^{(2)}$ coincide with $M^{(1)}, M^{(2)}$
if the dependence of $\tilde\psi_0$ on $\langle N_0\rangle$ is negligible
or vanishes, as e.g. in homogeneous systems.

$M^{(1)}_{\kappa,\nu\mu}$ and similarly $M'^{(1)}_{\kappa,\nu\mu}$
describes a scattering process
in which one atom is scattered out of the condensate by the absorption
of the two quasiparticles $\nu, \mu$ from - and the emission
of the new quasiparticle $\kappa$ into - the thermal bath. Likewise
$M^{(2)}_{\nu\mu,\kappa}$ and similarly $M'^{(2)}_{\nu\mu,\kappa}$
describes a scattering process where an incoming thermal 
quasiparticle
$\kappa$ is absorbed, again an atom is kicked out from the condensate, and two
quasiparticles $\nu, \mu$ are emitted into the thermal bath. The scattering amplitudes for both processes are linearly superposed due to the phase-coherence of the condensate which exists 
on the time-scale of the relaxation process induced by the scattering process
even if it is destroyed on a much longer time scale.

We can now calculate the correlation functions of the fluctuating forces.
Their averages over the bath of quasi-particles vanish,  $\langle\Re(\hat{\xi}(t))\rangle=0=\langle\Im(\hat{\xi}(t))\rangle$. Their second-order correlation functions
are obtained as
\begin{eqnarray}
\langle\Re(\hat{\xi}'(t))\Re(\hat{\xi}'(t'))\rangle_\phi &=& \frac{1}{8}\sum_{\kappa,\nu,\mu}|\big((M'^{(1)}_{\kappa,\nu\mu})^*+M'^{(2)}_{\nu\mu,\kappa}\big)|^2\nonumber\\
&& \qquad\qquad\Bigg\{\bar{n}_\kappa(\bar{n}_\nu+1)(\bar{n}_\mu+1)e^{i(\tilde\omega_\kappa-\tilde\omega_\nu-\tilde\omega_\mu)(t-t')}\nonumber\\
&&
\qquad\qquad+\bar{n}_\nu\bar{n}_\mu(\bar{n}_\kappa+1)e^{-i(\tilde\omega_\kappa-\tilde\omega_\nu-\tilde\omega_\mu)(t-t')}\Bigg\}\label{5.0}
\\
\langle\Im(\hat{\xi}(t))\Im(\hat{\xi}(t'))\rangle_\phi &=& \frac{1}{8}\sum_{\kappa,\nu,\mu}|(M^{(1)}_{\kappa,\nu\mu})^*-M^{(2)}_{\nu\mu,\kappa}|^2\nonumber\\
&& \qquad\qquad\Bigg\{\bar{n}_\kappa(\bar{n}_\nu+1)(\bar{n}_\mu+1)e^{i(\tilde\omega_\kappa-\tilde\omega_\nu-\tilde\omega_\mu)(t-t')}\nonumber\\
&&\qquad\qquad+\bar{n}_\nu\bar{n}_\mu(\bar{n}_\kappa+1)e^{-i(\tilde\omega_\kappa-\tilde\omega_\nu-\tilde\omega_\mu)(t-t')}\Bigg\}\label{5.1}\\
\langle\Re(\hat{\xi}'(t))\Im(\hat{\xi}(t'))\rangle_\phi &=& \frac{1}{8i}\sum_{\kappa,\nu,\mu}\Bigg\{\left((M_{\kappa,\nu\mu}^{(1)})^*-M_{\nu\mu,\kappa}^{(2)}\right)
\left(M'^{(1)}_{\kappa,\nu\mu}+(M'^{(2)}_{\nu\mu,\kappa})^*\right)\nonumber\\
&& 
\qquad\qquad\qquad(\bar{n}_\nu+1)(\bar{n}_\mu+1)\bar{n}_\kappa e^{i(\tilde\omega_\kappa-\tilde\omega_\nu-\tilde\omega_\mu)(t-t')}\nonumber\\
\label{5.2}\\ 
&& 
\qquad\qquad-\left(M_{\kappa,\nu\mu}^{(1)}-(M_{\nu\mu,\kappa}^{(2)})^*\right)
\left((M'^{(1)}_{\kappa,\nu\mu})^*+M'^{(2)}_{\nu\mu,\kappa}\right)\nonumber\\
&&
\qquad\qquad\qquad(\bar{n}_\kappa+1)\bar{n}_\nu\bar{n}_\mu e^{-i(\tilde\omega_\kappa-\tilde\omega_\nu-\tilde\omega_\mu)(t-t')}\Bigg\}\nonumber
\end{eqnarray}
These correlation functions can be replaced by delta-functions provided that the frequency-sums contain a flat quasi-continuum of nearly
resonant terms in a neighborhood of the resonance $\tilde\omega_\kappa-\tilde\omega_\nu-\tilde\omega_\mu=0$ which is broad compared to the damping rates we calculate here. This assumption will be satisfied in sufficiently large condensates. The strengths of the delta-functions can then be extracted from the expressions (\ref{5.0}), (\ref{5.1}), (\ref{5.2}) by taking the time-averages $
\int_{-\infty}^{\infty}d(t-t')\langle\hat{\xi}(t)\hat{\xi}(t')\rangle_\phi$
and
$\int_{-\infty}^{\infty}d(t-t')\langle\hat{\xi}^+(t)\hat{\xi}(t')\rangle_\phi
$.

$\Re(\hat{\xi}'(t)), \Im(\hat\xi(t))$ are here given as  expressions involving {\it operators}. Provided the Markovian approximation is satisfied  the average of their
  commutators   over the quasi-particle bath 
are again given by delta functions in time.
Explicitly we obtain for the coefficients of the delta-functions
\begin{eqnarray}
\int_{-\infty}^{\infty}dt\langle[\Re(\hat{\xi}'(t)),\Re(\hat{\xi}'(0))]\rangle_\phi
=0= 
&&\int_{-\infty}^{\infty}dt\langle[\Im(\hat\xi(t)),\Im(\hat{\xi}(0))]
\rangle_\phi\nonumber\\
\int_{-\infty}^{\infty}dt\langle[\Re(\hat{\xi'}(t)),\Im(\hat{\xi}(0))]
\rangle_\phi
=\frac{\pi}{8i}\sum_{\kappa,\nu,\mu}&&\Bigg(
\left((M_{\kappa,\nu\mu}^{(1)})^*-M_{\nu\mu,\kappa}^{(2)}\right)
\left(M'^{(1)}_{\kappa,\nu\mu}+(M'^{(2)}_{\nu\mu,\kappa})^*\right)-c.c\Bigg)
\label{comm}\\
&&\{(\bar{n}_\nu+1)(\bar{n}_\mu+1)\bar{n}_\kappa-(\bar{n}_\kappa+1)\bar{n}_\nu\bar{n}_\mu\}\delta(\tilde\omega_\kappa-\tilde\omega_\nu-\tilde\omega_\mu)
=0\nonumber
\end{eqnarray}
It can easily be verified that the bracket $\{...\}$ 
in the last line of (\ref{comm}) vanishes if it is multiplied by the delta-function expressing energy-conservation.
As a result the fluctuating force $\hat{\xi}$ in the Markoffian limit can indeed be treated as a c-number and will henceforth again be denoted by $\xi$. This also serves as a nice consistency-check that it is indeed possible to treat the condensate classically, even after taking its interaction with the quasi-particles into account. 

Let us now proceed to derive formulas for the three transport parameters $\Gamma_0, \Gamma'$ and $\Gamma''$. From 
\begin{eqnarray}
2\Gamma_0+\Gamma'=&&\frac{1}{\hbar k_BT}\int_{-\infty}^{+\infty} dt\langle \xi^*(t)\xi(0)\rangle_\phi\nonumber\\
\Gamma'+i\Gamma''=&&\frac{1}{\hbar k_BT}\int_{-\infty}^{+\infty} dt\langle \xi(t)\xi(0)\rangle_\phi
\end{eqnarray}
implied by eqs.(\ref{C1},\ref{C2}) we obtain the representations
\begin{eqnarray}
\Gamma_0=&&\frac{1}{\hbar k_BT}\int_\infty^\infty d(t-t')\langle\Im(\xi(t))\Im(\xi(t'))\rangle_\phi
\label{R2'}\\
\Gamma_0+\Gamma'=&&\frac{1}{\hbar k_BT}\int_\infty^\infty d(t-t')\langle\Re(\xi'(t))\Re(\xi'(t'))\rangle_\phi
\label{R1'}\\
\Gamma''=&&\frac{2}{\hbar k_BT }\int_\infty^\infty d(t-t')\langle\Re(\xi'(t))\Im(\xi(t'))\rangle_\phi.
\label{R3'}
\end{eqnarray}
which have the form of Green-Kubo relations for the transport coefficients.
Using the explicit form (\ref{Re},\ref{Im}) of the fluctuating forces the thermal 
averages can be taken and the time-integrals in eqs.(\ref{R1'},\ref{R2'},\ref{R3'}) can be carried out
which leads to the formulas
\begin{eqnarray}
\Gamma_0=\frac{\pi}{2\hbar k_BT}\sum_{\kappa,\nu,\mu}&&|(M^{(1)}_{\kappa,\nu\mu})^*-M^{(2)}_{\nu\mu,\kappa}|^2\bar{n}_\nu\bar{n}_\mu(\bar{n}_\kappa+1)\delta(\tilde\omega_\kappa-\tilde\omega_\mu-\tilde\omega_\nu)\label{34}\\
\Gamma_0+\Gamma'=\frac{\pi}{2\hbar k_BT}\sum_{\kappa,\nu,\mu}&&|(M'^{(1)}_{\kappa,\nu\mu})^*+M'^{(2)}_{\nu\mu,\kappa}|^2\bar{n}_\nu\bar{n}_\mu(\bar{n}_\kappa+1)\delta(\tilde\omega_\kappa-\tilde\omega_\mu-\tilde\omega_\nu)\label{31}\\
\Gamma''=\frac{-i\pi}{2\hbar k_BT}\sum_{\kappa,\nu,\mu}
&&\Bigg\{\left((M_{\kappa,\nu\mu}^{(1)})^*-M^{(2)}_{\nu\mu,\kappa}\right)
\left(M'^{(1)}_{\kappa,\nu\mu}+(M'^{(2)}_{\nu\mu,\kappa})^*\right)\nonumber\\
&& 
-\left(M_{\kappa,\nu\mu}^{(1)}-(M_{\nu\mu,\kappa}^{(2)})^*\right)
\left((M'^{(1)}_{\kappa,\nu\mu})^*+M'^{(2)}_{\nu\mu,\kappa}\right)\Bigg\}
\label{31a}\\
&&\bar{n}_\nu\bar{n}_\mu(\bar{n}_\kappa+1)\delta(\tilde\omega_\kappa-\tilde\omega_\mu-\tilde\omega_\nu)
\nonumber
\end{eqnarray}
These expressions
constitute our general results for the three transport parameters.
They have to be evaluated separately
for each individual trap geometry.

\section{Relation to the fluctuation and dissipation of the excitations}\label{sec.5a}

As was pointed out after eq.(\ref{26b}) by phenomenological arguments, the noise term $\Re(\xi(t))$ is not connected with the fluctuations of the number of particles in the condensate, but must be due to other fluctuations, which are then necessarily thermal fluctuations of the amplitudes of the excited states. In our microscopic results this can be seen from the fact that the fluctuating force $\Re(\hat\xi'(t))$ according to eq.(\ref{59a}) contains precisely the same operator which also appears in $\hat H_3(t)$
and couples the atoms in the thermal cloud to the condensate. 

In the special case where the difference between the coupling matrix-elements 
$M'^{(1,2)}$ and $M^{(1,2)}$ is negligible (which is exactly satisfied in box-like traps, cf.section \ref{sec:6})
 the intensity $\Gamma_0+\Gamma^{\prime}$ of the noise source can  be expressed 
entirely as a property of the excitations, as we shall now demonstrate.\footnote{In the general case the coupling of the condensate to the non-condensate modes differs from the coupling between the non-condensate modes and the relation between $\Gamma_0+\Gamma'$ and the $\gamma_\nu$ is less direct.}  For their amplitudes $\hat{\alpha}_{\nu}(t), \hat{\alpha}^+_{\nu}(t)$ a quantum-Langevin equation can be derived microscopically along the same lines employed here for the condensate amplitude. We have done this elsewhere \cite{Nic}, see also \cite{PRL} with the result, in the Markoffian limit,
\begin{eqnarray}
   \frac{d\hat{\alpha}_{\nu}(t)}{dt}=
       - i\omega_{\nu}\hat{\alpha}_{\nu}(t) - \gamma_{\nu}
\hat{\alpha}_{\nu}(t)
       + \hat{\xi}_\nu(t)
\end{eqnarray}
with Gaussian fluctuating force-operators with vanishing average and
\begin{eqnarray}
       \langle\hat{\xi}^+_{\nu}(t)\hat{\xi}_{\mu}(t^{\prime})\rangle
     &&= 2\gamma_{\nu}\bar{n}_{\nu}\delta(t-t^{\prime})\delta_{\nu\mu}\nonumber\\
\label{xy}\\
       \langle[\hat{\xi}_{\nu}(t),\hat{\xi}^+_{\mu}(t^{\prime})]\rangle
     &&= 2\gamma_{\nu}\delta(t-t^{\prime})\delta_{\nu\mu}\nonumber
\end{eqnarray}
where the damping rates $\gamma_{\nu}$ are given by
\begin{eqnarray}
\gamma_\nu=\frac{\pi \langle N_0\rangle}{\hbar^2}\sum_{\kappa,\mu}\big \{&&|(M_{\kappa,\nu\mu}^{(1)})^*+M_{\nu\mu,\kappa}^{(2)}|^2(\bar{n}_\mu-\bar{n}_\kappa)\delta(\omega_\kappa-\omega_\mu-\omega_\nu)\nonumber\\
&&+|M_{\nu,\kappa\mu}^{(1)}+(M_{\kappa\mu,\nu}^{(2)})^*|^2(\bar{n}_\kappa+\frac{1}{2})\delta(\omega_\kappa+\omega_\mu-\omega_\nu)\big\}.
\label{gammanu}
\end{eqnarray}
The first term describes Landau-damping of the mode $\nu$ by scattering a quasi-particle from mode $\mu$ to mode $\kappa$ and is equivalent to a result derived in \cite{Pitaevskii} by the golden rule.
The second term in eq.(\ref{gammanu}) describes Beliaev damping, where the mode $\nu$ decays into two modes $\kappa, \mu$. It survives even for $T\rightarrow 0$ where $\bar{n}_\kappa \rightarrow 0$ for all modes.

Let us now establish the connection between $\Gamma_0+\Gamma^{\prime}$ and the damping rates $\gamma_{\nu}$ as given by (\ref{gammanu}). We shall show that the simple sum-rule
\begin{eqnarray}
      \Gamma_0+\Gamma^{\prime}
    = \frac{\hbar}{3\langle N_0\rangle k_BT}
            \sum\limits_{\nu}\bar{n}_{\nu}(\bar{n}_{\nu}+1)\gamma_{\nu}
\end{eqnarray}
holds.
To see this we need to consider
\begin{eqnarray}
       \sum\limits_{\nu}\bar{n}_{\nu}(\bar{n}_{\nu}+1)\gamma_{\nu}
    &=& \frac{\pi\langle N_0\rangle}{\hbar^2}
       \sum\limits_{\kappa\mu\nu} |(M^{(1)}_{\kappa,\nu\mu})^*
     +  M^{(2)}_{\nu\mu,\kappa}|^2\delta(\omega_{\kappa}-\omega_{\nu}
        -\omega_{\mu}) \nonumber\\
       && \cdot
        \Big\{
        (\bar{n}_{\mu}-\bar{n}_{\kappa})\bar{n}_{\nu}(\bar{n}_{\nu}+1)
      +  \frac{1}{2}(\bar{n}_{\mu}+\bar{n}_{\nu}+1)
              \bar{n}_{\kappa}(\bar{n}_{\kappa}+1)\Big\}
 \label{eq:K1}
\end{eqnarray}
The second term in the curly bracket arises from the second term in (\ref{gammanu}) by first exchanging the notations for the summation indices
$\nu$ and $\kappa$ and then symmetrizing in $\nu$ and $\mu$, because the matrix-elements are already symmetric in these indices. 
The remainder of the proof then consists simply in noting that 
for $\omega_{\kappa} = \omega_{\nu}+\omega_{\mu}$ the identities
\begin{eqnarray}
       (\bar{n}_{\mu}-\bar{n}_{\kappa})\bar{n}_{\nu}(\bar{n}_{\nu}+1)
     &&= \bar{n}_{\mu}\bar{n}_{\nu}(\bar{n}_{\kappa}+1)\nonumber\\
\label{rel}\\
       (\bar{n}_{\mu}+\bar{n}_{\nu}+1)\bar{n}_{\kappa}(\bar{n}_{\kappa}+1)
     &&= \bar{n}_{\mu}\bar{n}_{\nu}(\bar{n}_{\kappa}+1)\nonumber
\end{eqnarray}
hold. Using this in eq.(\ref{eq:K1}) and then comparing with (\ref{31}) establishes the sum-rule. We can also
note that the processes due to Landau scattering contribute to
the sum-rule with precisely twice the strength of those due
to Beliaev scattering.

Thus we see that, in general, the noise-amplitudes proportional to the combination of matrix-elements $(M^{(1)}_{\kappa,\nu\mu})^*-M^{(2)}_{\nu\mu,\kappa}$ drive the {\it number}-fluctuations in the condensate, while those proportional to $(M'^{(1)}_{\kappa,\nu\mu})^*+M'^{(2)}_{\nu\mu,\kappa}$ are due to fluctuations of the occupation numbers in the excited states, couple in the Hamiltonian
to the
particle number in the condensate and therefore  drive the {\it phase}-fluctuations in the condensate.

\section{Evaluation of the Transport Parameters for a Box-like
         Trap}\label{sec:6}

For simplicity we consider now a trap consisting of a cube of length $L$ with cyclic boundary conditions. In the following equilibrium values of all parameters are implied, but we shall omit in this section the tilde  and write $\mu$ for $\langle\mu\rangle$ to simplify our notation. The normalized $u$ and $v$ coefficients are in this case

\begin{eqnarray}
        u_\nu
    &=& \frac{E_{\nu}+p^2_{\nu}/2m}{\sqrt{2E_{\nu}p^2_{\nu}/m}}\,\,
        \frac{1}{\sqrt{V}}\,\,e^{i{\vec{p}_{\nu}}\cdot \vec{x}/\hbar}
\end{eqnarray}
\begin{eqnarray}
        v_{\nu}
    &=& - \frac{E_{\nu}-p^2_{\nu}/2m}{\sqrt{2E_{\nu}p^2_{\nu}/m}}\,\,
        \frac{1}{\sqrt{V}}\,\,e^{i{\vec{p}_{\nu}}\cdot \vec{x}/\hbar}
\end{eqnarray}
with
\begin{eqnarray}
        E_{\nu} =
       \sqrt{\left(\frac{p^2_{\nu}}{2m}+\mu\right)^2-\mu^2}
\end{eqnarray}
and $\vec{p}_{\nu} = \hbar\frac{2\pi}{L} \vec{n}_{\nu}$ with integer vector $\vec{n}_{\nu}$.

\subsection{Transport coefficients}\label{sub:a}

The squares of the relevant  matrix elements for $E_{\kappa} = E_{\nu}+E_{\mu}$ become
\begin{eqnarray}
 \left|\left(M^{(1)}_{\kappa,\nu\mu}\right)^*-M^{(2)}_{\nu\mu,\kappa}\right|^2
    &&= \left(\frac{U_0}{V}\right)^2
       \frac{G(E_\nu,E_\mu,\mu)}{E_{\nu}E_{\mu}E_{\kappa}}        \delta_{\vec{n}_{\kappa}, \vec{n}_{\nu}+\vec{n}_{\mu}}\\
\left|\left(M^{(1)}_{\kappa,\nu\mu}\right)^*+M^{(2)}_{\nu\mu,\kappa}\right|^2
     &&= \left(\frac{U_0}{V}\right)^2
       \frac{G(E_\nu,E_\mu,-\mu)}{E_{\nu}E_{\mu}E_{\kappa}} 
       \delta_{\vec{n}_{\kappa}, \vec{n}_{\nu}+\vec{n}_{\mu}}
\end{eqnarray}
with
\begin{eqnarray}
G(x,y,\alpha)=&&\sqrt{\alpha^2+(x+y)^2}\left(3\sqrt{\alpha^2+x^2}\sqrt{\alpha^2+y^2}-xy+2\alpha(\sqrt{\alpha^2+x^2}+\sqrt{\alpha^2+y^2}+\alpha)\right)\nonumber\\
&&+(x+y)\left(x(\sqrt{\alpha^2+y^2}+\alpha)+y(\sqrt{\alpha^2+x^2}+\alpha)\right)\label{G}\\
&&+2\alpha(\sqrt{\alpha^2+x^2}+\alpha)(\sqrt{\alpha^2+y^2}+\alpha)+\alpha(x^2+y^2+\alpha^2)\nonumber
\end{eqnarray}
The transport coefficients are then expressed as
the simple result
\begin{eqnarray}
       \Gamma^"=0
\end{eqnarray}
and
\begin{eqnarray}
       \Gamma_0
     &&= \frac{2}{\pi}\left(\frac{a}{L}\right)^2
       \sum_{{\vec{n}_{\nu}},\vec{n}_{\mu}}
    \frac{\delta (\varepsilon_{\nu}+\varepsilon_{\mu}-
            \varepsilon_{\nu+\mu})}{\varepsilon_{\nu}\varepsilon_{\mu}
            (\varepsilon_{\nu}+\varepsilon_{\mu})}
    G(\varepsilon_\nu,\varepsilon_\mu,\alpha)       F(\varepsilon_{\nu},\varepsilon_{\mu},\frac
       {k_BT}{\hbar\omega_0})\label{dummy}\\
\nonumber\\
       \Gamma_0+\Gamma^{\prime}&&=\frac{2}{\pi}\left(\frac{a}{L}\right)^2
       \sum_{{\vec{n}_{\nu}},\vec{n}_{\mu}}
       \frac{\delta (\varepsilon_{\nu}+\varepsilon_{\mu}-\varepsilon_{\nu+\mu})}
      {\varepsilon_{\nu}\varepsilon_{\mu}(\varepsilon_{\nu}+\varepsilon_{\mu})}
        G(\varepsilon_\nu,\varepsilon_\mu,-\alpha) F(\varepsilon_{\nu},\varepsilon_{\mu},
       \frac{k_BT}{\hbar\omega_0})\label{dummy2}
\end{eqnarray}
with
\begin{eqnarray}
       F(\varepsilon_{\nu},\varepsilon_{\mu},\frac{k_BT}
       {\hbar\omega_0})=\frac{\hbar\omega_0}{k_BT}
      \frac{e^{\beta\hbar\omega_0(\varepsilon_{\nu}+\varepsilon_{\mu})}}
       {\left(e^{\beta\hbar\omega_0(\varepsilon_{\nu}+\varepsilon_{\mu})}
              -1\right)
        \left(e^{\beta\hbar\omega_0\varepsilon_{\nu}}-1\right)
        \left(e^{\beta\hbar\omega_0\varepsilon_{\mu}}-1\right)}
\end{eqnarray}
Here we scaled the scattering length $a=mU_0/4\pi\hbar^2$ with $L$ and the energies $E_{\nu}, E_{\mu}$ and $\mu, k_BT$ with the energy $\hbar\omega_0=(2\pi\hbar)^2/2mL^2$, defining
\begin{eqnarray}
       \varepsilon_{\nu}=
       \sqrt{(n^2_{\nu}+\alpha)^2-\alpha^2}
\end{eqnarray}
\begin{eqnarray}
       \varepsilon_{\nu+\mu}=
       \sqrt{\left( (\vec{n}_{\nu}+\vec{n}_{\mu})^2
       + \alpha\right)^2 - \alpha^2}
\end{eqnarray}
with $\alpha=\mu/\hbar\omega_0.$

The double sums over $\vec{n}_{\nu}, \vec{n}_{\mu}$ start with $\vec{n}$-values with $|\vec{n}|=1$. They are approximated by integrals according to
\begin{eqnarray}
       \sum_{\vec{n}_{\nu},\vec{n}_{\mu}}
       \delta\left(\varepsilon_{\nu}+\varepsilon_{\mu}-\varepsilon_{\nu+\mu}
       \right)(...)
     = \pi^2\int\limits^{\infty}_{\sqrt{1+2\alpha}} 
       \int\limits^{\infty}_{\sqrt{1+2\alpha}}
        \frac{\varepsilon_{\nu}\varepsilon_{\mu}
 (\varepsilon_{\nu}+\varepsilon_{\mu})
(...)d\varepsilon_{\nu}d\varepsilon_{\mu}}
       {\sqrt{(\varepsilon^2_{\nu}+\alpha^2)
       (\varepsilon^2_{\mu}+\alpha^2)
       ((\varepsilon_{\nu}+\varepsilon_{\mu})^2+\alpha^2)}}
\end{eqnarray}
Here (...) is any smooth function of $\varepsilon_{\nu}, \varepsilon_{\mu}$. In all experiments so far $\alpha\gg1$ is satisfied, i.e. we can replace $\sqrt{1+2 \alpha}\rightarrow \sqrt{2\alpha}$.
This leaves us with the integral expressions
\begin{eqnarray}
       \Gamma_0&&=2\pi\Big(\frac{a}{L}\Big)^2
                    \int\limits^{\infty}_{\sqrt{2\alpha}}
       \int\limits^{\infty}_{\sqrt{2\alpha}}
    \frac{  G(\varepsilon_\nu,\varepsilon_\mu,\alpha) F(\varepsilon_{\nu},\varepsilon_{\mu},
             \frac{k_BT}{\hbar\omega_0})d\varepsilon_{\nu}d\varepsilon_{\mu}   }
       {    \sqrt{(\varepsilon^2_{\nu}+\alpha^2)
       (\varepsilon^2_{\mu}+\alpha^2)
       ((\varepsilon_{\nu}+\varepsilon_{\mu})^2+\alpha^2)}  }\label{dummy3}\\
\nonumber\\
       \Gamma_0 +\Gamma^{\prime} &&= 2\pi\Big(\frac{a}{L}\Big)^2
                    \int\limits^{\infty}_{\sqrt{2\alpha}}
       \int\limits^{\infty}_{\sqrt{2\alpha}}
    \frac{  G(\varepsilon_\nu,\varepsilon_\mu,-\alpha) F (\varepsilon_{\nu},\varepsilon_{\mu},
            \frac{k_BT}{\hbar\omega_0})d\varepsilon_{\nu}d\varepsilon_{\mu}  }
            {  \sqrt{(\varepsilon^2_{\nu}+\alpha^2)
       (\varepsilon^2_{\mu}+\alpha^2)
       ((\varepsilon_{\nu}+\varepsilon_{\mu})^2+\alpha^2)} }
   \label{dummy4}
\end{eqnarray}
The expression for $\Gamma_0$ and the asymptotic behavior 
for $\varepsilon_\nu, \varepsilon_\mu\rightarrow 0$: $G(\varepsilon_\nu,\varepsilon_\mu,|\alpha|) F(\varepsilon_{\nu},\varepsilon_{\mu},k_BT/\hbar\omega_0)\rightarrow 18(k_BT/\hbar\omega_0)^21/[\varepsilon_\nu\varepsilon_\mu(\varepsilon_\nu+\varepsilon_\mu)]$ make it amply clear that the states with the smallest energies $\varepsilon_{\nu}\ll\alpha$ make a large contribution to $\Gamma_0$ (but {\it not} to $\Gamma_0+\Gamma'$). To calculate this contribution it is permitted \footnote{We actually need the additional condition $\sqrt{\mu\hbar\omega_0} \ll k_BT$} to use $\beta\hbar\omega_0\varepsilon_{\nu}\ll 1$, $\beta\hbar\omega_0\varepsilon_{\mu}\ll 1$ under the integral and to approximate
\begin{eqnarray}
       F(\varepsilon_{\nu},\varepsilon_{\mu},
          \frac{k_BT}{\hbar\omega_0})
        = \left(\frac{k_BT}{\hbar\omega_0}\right)^2
        \frac{1}{(\varepsilon_{\nu}+\varepsilon{_\mu})
                   \varepsilon_{\nu}\varepsilon{_\mu}}\label{Fapp}
\end{eqnarray}
in addition to approximating $\sqrt{\varepsilon^2_{\nu}+\alpha^2} \sqrt{\varepsilon^2_{\mu}+\alpha^2} \sqrt{(\varepsilon_{\nu}+\varepsilon_{\mu})^2+\alpha^2} \cong\alpha^3$, and neglecting terms of order $\varepsilon^2_{\nu}/\alpha^2$.

This contribution to $\Gamma_0$, which we shall denote as $\Gamma_{00}$, then reduces to
\begin{eqnarray}
      \Gamma_{00}=36\pi\left(\frac{a}{L}\right)^2
             \left(\frac{k_BT}{\hbar\omega_0}\right)^2
           \int\limits^{\infty}_{\sqrt{2\alpha}}d\varepsilon_{\nu}
           \int\limits^{\infty}_{\sqrt{2\alpha}}d\varepsilon_{\mu}
           \frac{1}{\varepsilon_{\nu}\varepsilon{_\mu}
                   (\varepsilon_{\nu}+\varepsilon{_\mu})}\label{G0}
\end{eqnarray}
The double-integral can be evaluated as $\sqrt{2/\alpha} \ln2$ which yields the final result
\begin{eqnarray}
      \Gamma_{00}=36\pi\sqrt{2}\ln2 \left(\frac{a}{L}\right)^2
                 \left(\frac{\hbar\omega_0}{\mu}\right)^{1/2}
 \left(\frac{k_BT}{\hbar\omega_0}\right)^2=1.59..\left(\frac{T}{T_c}\right)^2\left(\frac{N}{\langle N_0\rangle}\right)^{1/2}N^{1/3}\left(\frac{k_BT_ca^2m}{\hbar^2}\right)^{3/4}\label{104}\label{G00}
\end{eqnarray}
The second form of this expression is obtained by eliminating $V=L^3$ in favor of the critical temperature of the equivalent ideal Bose-gas at the same density via \begin{equation}
V=(N/\zeta(3/2))(2\pi\hbar^2/k_BT_cm)^{3/2}.\label{Vol}
\end{equation}

There is yet another particularly important contribution to $\Gamma_0$ due to the infrared-singularity of the integrand, we shall denote it as $\Gamma_{01}$, where only {\it one} of the two excitation frequencies, say $E_\nu$, is small compared to $\mu$,
while the other is larger, of order $\mu$ or even $k_BT$. With respect to $\epsilon_\nu$ the low-energy asymptotics may then still be used.
We then obtain
\begin{equation}
\Gamma_{01}=\frac{4\pi}{\alpha}\left(\frac{a}{L}\right)^2\int\limits^{q\alpha}_{\sqrt{2\alpha}}d\varepsilon_\nu\int\limits^{\infty}_{\sqrt{2\alpha}}d\varepsilon_\mu
\frac{G(0,\varepsilon_\mu,\alpha)-G(0,0,\alpha)}{\varepsilon_\nu(\varepsilon_\mu^2+\alpha^2)}\frac{e^{\beta\hbar\omega_0\varepsilon_\mu}}{(e^{\beta\hbar\omega_0\varepsilon_\mu}-1)^2},
\end{equation}
where we have set an upper cut-off for the small energy at a fraction $q$ of the chemical potential. Most of the $\varepsilon_\mu$-integral comes from
a range around $\alpha$ and we may therefore replace the thermal function
by its asymptotics for $\beta\hbar\omega_0\varepsilon_\mu\rightarrow 0$, which is $(k_BT/\hbar\omega_0\varepsilon_\mu)^{2}$.
The integrals can then be performed with the result
\begin{equation}
\Gamma_{01}=16\pi\left(\frac{a}{L}\right)^2
\ln{\left(\frac{q^2\mu}{2\hbar\omega_0}\right)}
\frac{(k_BT)^2}{\hbar\omega_0\mu}
=1.22..\left(\frac{T}{T_c}\right)^2\frac{N}{\langle N_0\rangle}\left(\frac{k_BT_ca^2m}{\hbar^2}\right)^{1/2}\ln\left(\frac{q^2\mu}{2\hbar\omega_0}\right).
\end{equation}
We  conclude that this contribution to $\Gamma_0$ is smaller than the leading term $\Gamma_{00}$ by the order of magnitude $(\hbar\omega_0/\mu)^{1/2}
\ln(q^2\mu/2\hbar\omega_0)$.

Let us now turn to the expressions for $\Gamma_0-\Gamma_{00}-\Gamma_{01}$  and $\Gamma_0+\Gamma^{\prime}$. They can be simplified for $\alpha\gg 1$ by rescaling $\varepsilon_{\nu}, \varepsilon_{\mu}$ by $\alpha$ and taking the limit $\sqrt{2/\alpha}\rightarrow 0$ for the lower boundaries of the rescaled integrals. We find in this way
\begin{eqnarray}
       \Gamma_0-\Gamma_{00}-\Gamma_{01}
       =        2\pi\Big(\frac{a}{L}\big)^2
                \Big(\frac{\mu}{\hbar\omega_0}\Big)
               \int\limits^{\infty}_0 dx  \int\limits^{\infty}_0 dy
          \Biggl[\frac{  G(x,y,1)  F(x,y,\frac{k_BT}{\mu}) }
                      {   \sqrt{(x^2+1)(y^2+1)((x+y)^2+1)   }}&&\nonumber\\ 
         -\Big(\frac{k_BT}{\mu}\Big)^2\Big(\frac{18}{xy(x+y)}&&\label{dummy5}\\
 +8 \frac{x(x^2+1)(y^2+\sqrt{y^2+1}-1)
+y(y^2+1)(x^2+\sqrt{x^2+1}-1)}{x^2y^2(x^2+1)(y^2+1)}&&\Big)\Biggl]\nonumber\\ 
\nonumber\\
       \Gamma_0+\Gamma^{\prime}=2\pi
              \Big(\frac{a}{L}\Big)^2
              \Big(\frac{\mu}{\hbar\omega_0}\Big)
      \int\limits^{\infty}_0 dx  \int\limits^{\infty}_0 dy 
    \frac{G(x,y,-1) F(x,y,\frac{k_BT}{\mu})}
{\sqrt{(x^2+1)(y^2+1)((x+y)^2+1) }}&&\label{dummy6}
   \end{eqnarray}
where we used the scaling property
\begin{eqnarray}
      F(\varepsilon_{\nu},\varepsilon_{\mu},
              \frac{k_BT}{\hbar\omega_0})
        =   \frac{1}{\alpha}  F(\frac{\varepsilon_{\nu}}{\alpha},
                     \frac{\varepsilon_{\mu}}{\alpha},
                      \frac{k_BT}{\hbar\omega_0\alpha} )
                      \end{eqnarray}
Eqs(\ref{dummy5},\ref{dummy6}) are the complete result for
the temperature dependent transport parameters of the condensate
for box-like traps. In general the integrals have to be done numerically. We shall here consider some asymptotic results only.

First we consider these expressions asymptotically for $k_BT/\mu\gg1$. Then the integrals receive important contributions from x, y of the order of 1, i.e. from quasi-particle-energies of the order of the chemical potential, and also from values of x,y large compared to 1, i.e. quasi-particle energies of order $k_BT$. The contributions $\Gamma_0^{(>)}$ and $ \Gamma'^{(>)}$ from large energies can be determined in leading power in $(k_BT/\mu)$  by approximating
\begin{eqnarray}
       F(x,y,\frac{k_BT}{\mu})\simeq\frac{\mu}{k_BT}
          e^{-\frac{\mu}{k_BT}(x+y)}
\label{eq:app}
\end{eqnarray}
and rescaling $x$ and $y$ by $k_BT/\mu$. In the integrals\footnote{In the high-energy regime the subtractions of the infrared-divergent terms
in the integrand of (\ref{dummy5}) are of no importance.} for $\Gamma_0$ and $\Gamma_0+\Gamma'$ we can then let $\mu/k_BT\rightarrow0$ without any problem, using the property $G(x,y,0)=4xy(x+y)$.whereupon they are easily evaluated with the asymptotic results
\begin{eqnarray}
      \Gamma_0^{(>)}+\Gamma'^{(>)}\simeq \Gamma_0^{(>)}\simeq8\pi \left(\frac{a}{L}\right)^2
             \frac{k_BT}{\hbar\omega_0}
=1.27..\frac{T}{T_c}\frac{k_BT_ca^2m}{\hbar^2}\label{Diff}
\end{eqnarray}
We see that the result for $\Gamma'^{(>)}$ vanishes to this order.

For the contributions $\Gamma_0^{(\mu)},\Gamma'^{(\mu)}$ from quasi-particles with energies around $\mu$ we can approximate $F(x,y,k_BT/\mu)$ according to eq.(\ref{Fapp})
and find
\begin{eqnarray}
       \Gamma_0^{(\mu)}&&\simeq B_0^{(\mu)}\left(\frac{a}{L}\right)^2
        \frac{(k_BT)^2}{\hbar\omega_0\mu}=0.0243..B_0^{(\mu)}\left(\frac{T}{T_c}\right)^2\frac{N}{\langle N_0\rangle}\left(\frac{k_BT_ca^2m}{\hbar^2}\right)^{1/2}\label{Diff1}\\
     \Gamma_0^{(\mu)}+\Gamma'^{(\mu)}&&\simeq B'^{(\mu)}\left(\frac{a}{L}\right)^2
 \frac{(k_BT)^2}{\hbar\omega_0\mu}=0.0243..B'^{(\mu)}\left(\frac{T}{T_c}\right)^2\frac{N}{\langle N_0\rangle}\left(\frac{k_BT_ca^2m}{\hbar^2}\right)^{1/2}\label{endl}\end{eqnarray}
with the numbers $B_0^{(\mu)}, B'^{(\mu)}$ defined by the integrals
\begin{eqnarray}
B_0^{(\mu)}=2\pi\int\limits^{\infty}_0 dx  \int\limits^{\infty}_0 dy 
    &&\Big[\frac{G(x,y,1)}
{\sqrt{(x^2+1)(y^2+1)((x+y)^2+1)}xy(x+y)}\nonumber\\
&&-\frac{18}{xy(x+y)}\\
&&-8 \frac{x(x^2+1)(y^2+\sqrt{y^2+1}-1)
+y(y^2+1)(x^2+\sqrt{x^2+1}-1)}{x^2y^2(x^2+1)(y^2+1)}\Big]\nonumber\\
B'^{(\mu)}=2\pi\int\limits^{\infty}_0 dx  \int\limits^{\infty}_0 dy 
    &&\frac{G(x,y,-1)}
{\sqrt{(x^2+1)(y^2+1)((x+y)^2+1) }xy(x+y)}
\end{eqnarray}
We can conclude that the contribution from quasi-particles at energies of order $\mu$ is larger (for $\Gamma_0$ by an order of magnitude $k_BT/\mu$) than
the contribution from energies of order $k_BT$, but $\Gamma_0^{(\mu)}$
is, in large condensates, still subdominant to $\Gamma_{00}$ by the order of magnitude $\sqrt{\hbar\omega_0/\mu}$.

Now let us consider also the low-temperature limit, namely $k_BT/\mu\ll1$ or, equivalently, $T/T_c\ll\left(k_BT_ca^2m/\hbar^2\right)^{1/2}$. In this region it is not neccessary to distinguish $N$ and $\langle N_0\rangle$. The integrals now receive their contributions for $x, y$ both small compared to 1, but we can still use the approximation (\ref{eq:app}). 
For small $x, y$ we can expand
\begin{eqnarray}
G(x,y,1)\simeq18+3(x^2+y^2+xy)\qquad\qquad G(x,y,-1)\simeq \frac{9}{32}x^2y^2(x+y)^2\end{eqnarray}
To obtain the leading term it is enough to keep only the smallest powers of $x, y$ in the integrands. The integrals are easily evaluated with the asymptotic low-temperature results
\begin{eqnarray}
       \Gamma_0-\Gamma_{00}
       =  36\pi\left(\frac{a}{L}\right)^2
  \frac{k_BT}{\hbar\omega_0}=5.72..\frac{T}{T_c}\frac{k_BT_ca^2m}{\hbar^2}
\end{eqnarray}
\begin{eqnarray}
       \Gamma_0+\Gamma^{\prime}
       =   \frac{189}{2}\pi
              \left(\frac{a}{L}\right)^2\frac{\mu}{\hbar\omega_0}
 \left(\frac{k_BT}{\mu}\right)^7=0.366..\left(\frac{T}{T_c}\right)^7\left(\frac{k_BT_ca^2m}{\hbar^2}\right)^{-2}
\end{eqnarray}
As long as the temperature is high enough to satisfy $k_BT\gg\sqrt{\mu\hbar\omega_0}$ the part $\Gamma_{00}$  still dominates the value of $\Gamma_0$.

\subsection{Particle-number fluctuations}

We follow the procedure of Giorgini et al.\cite{11a} and deduce the particle-number fluctuations in the condensate from the number fluctuation in the thermal cloud. This leads to eq.(\ref{deltan}), which we evaluate using the expressions for $u_{\nu}, v_{\nu}$ and $E_{\nu}$.

We obtain
\begin{eqnarray}
       \langle\Delta N^2_0\rangle = \sum_{\nu}
     \frac{2\bar{n}_{\nu}(\bar{n}_\nu+1)(E^2_{\nu}+2\mu^2)+\mu^2}
             {2E^2_{\nu}}
\end{eqnarray}
Approximated by an integral this becomes
\begin{eqnarray}
       \langle\Delta N^2_0\rangle = \pi\int^{\infty}_{\sqrt{1+2\alpha}}
               \frac{d \varepsilon}{\varepsilon}
       \sqrt{
            \frac{\sqrt{\varepsilon^2+\alpha^2}-\alpha}
                 {\varepsilon^2+\alpha^2}                }
       \left(\alpha^2+
            \frac{\varepsilon^2+2\alpha^2}
                 {2(\sin h \frac{\beta\hbar\omega_0 \varepsilon}{2})^2}
       \right)
\end{eqnarray}
The dominant contribution comes from the lower boundary of the integration \cite{11a} which contributes, for $\alpha\gg 1$
\begin{eqnarray}
      \langle\Delta N^2_0\rangle \simeq 2\pi
      \left(\frac{k_BT}{\hbar\omega_0} \right)^2 =
      A^{\prime}\left(\frac{m k_BT}{\hbar^2}\right)^2 V^{4/3}\label{dom}
\end{eqnarray}
with
\begin{eqnarray}
       A^{\prime} = \frac{1}{2\pi^3} = 0.0161..
\end{eqnarray}
More precisely the dominant contribution to $\langle\Delta N^2_0\rangle$ is given by the discrete sum \cite{11a}
\begin{eqnarray}
       \langle\Delta N^2_0\rangle = 2\mu^2
               (k_BT)^2
            \sum_{\nu}\frac{1}{E^{4}_{\nu}}
\end{eqnarray}
which gives the same expression as (\ref{dom}) but with the prefactor\footnote{The formula (\ref{Mist1}) differs from the one given in \cite{11a} by a factor $2^{-4}$ whereas the numerical result differs by yet another factor; the formula (\ref{Mist2}) differs from the one in \cite{11a} by a factor 2.}
\begin{eqnarray}
       A =
       \frac{2}{(2\pi)^4}        \sum_{\vec{n}_{\nu}\not= 0}\,\,
       \frac{1}{n^4_{\nu}}   =   0.021..\label{Mist1}
\end{eqnarray}
If we eliminate the volume in favour of the critical temperature of the ideal Bose-gas of the same density via eq.(\ref{Vol}) we get
\begin{eqnarray}
       \langle\Delta N^2_0\rangle =
       A \frac{(2\pi)^2}{\zeta(3/2)^{4/3}} \left(\frac{T}{T_c}\right)^2 N^{4/3}.
\end{eqnarray}
At temperature $T=0$ a similar evaluation of (\ref{deltan}) gives
\begin{eqnarray}
      \langle\Delta N^2_0\rangle |_{T=0} = 2\sqrt{\pi} (aN)^{3/2}V^{-1/2}.\label{Mist2}
\end{eqnarray}

\subsection{Particle-number relaxation rate}

We are now in a position to evaluate the rate $\tau^{-1}_c$ from eq.(\ref{tau}) using the results for $\langle\Delta N^2_0\rangle$ (numbers are
calculated with the prefactor $A^{\prime}$) and $\Gamma_0\simeq\Gamma_{00}$. We obtain
\begin{eqnarray}
      \gamma_c = \frac{1}{\tau_c} = 18.0.. \frac{T}{T_c}\sqrt{\frac{\langle N_0\rangle}{N}}
               \left(\frac{k_BT_c a^2 m}{\hbar^2}\right)^{3/4}
               \frac{k_BT_c}{\hbar}
\end{eqnarray}
This result applies also in the low-temperature region, because it makes use only of the results for $\langle\Delta N^2_0\rangle$ and $\Gamma_0$ which also hold in that region.

To get an idea of order of magnitudes we compare this and the following results with the damping rate $\gamma_0$ of the lowest lying modes, which is given by \cite{Pitaevskii,Nic}
\begin{eqnarray}
      \gamma_0 = \frac{3\pi^2}{4} \left(\frac{a}{L}\right)\frac{k_BT}{\hbar}
               = 4.06.. \frac{T}{T_c} N^{-1/3}
                 \left(\frac{k_BT_c a^2m}{\hbar^2}\right)^{1/2}
                       \frac{k_BT_c}{\hbar}\label{gamma0}
\end{eqnarray}
We see that the relaxation rate $\gamma_c$ is of the order
\begin{eqnarray}
      \gamma_c  \sim
          N^{1/3}\left(\frac{k_BT_c a^2m}{\hbar^2}\right)^{1/4}\gamma_0
\end{eqnarray}
The proportionality factor is of the order of $\sqrt{\mu/\hbar\omega_0}$ and is large in large and strongly interacting condensates. I.e. then the relaxation of the condensate to its equilibrium is faster than the relaxation of the low-lying collective modes, but slower than the frequency of the lowest lying modes, which is $\sqrt{2\omega_0\mu/\hbar}$.

\subsection{Phase collapse rate}

The phase collapse rate is given by eq.(\ref{coll}) and requires only the result for $\langle\Delta\ N_0^2\rangle$ and $\Gamma''=0$. At zero temperature it reduces to
\begin{eqnarray}
       \gamma_{collapse}|_{T=0}
      = \frac{1}{\hbar}\frac{\partial\mu}{\partial \langle N_0\rangle}
        \sqrt{\langle\Delta N^2_0\rangle |_{T=0}}
\end{eqnarray}
from which we get
\begin{eqnarray}
       \gamma_{collapse}|_{T=0}
      = \frac{23.6..}{\sqrt{V}}(a n_0)^{3/4}\frac{\hbar a}{m}
      = 2.50..\frac{k_BT_c}{\hbar}
        \left(
              \frac{k_BT_c a^2m}{\hbar^2}        \right)^{7/8}
        \frac{1}{\sqrt N }
\end{eqnarray}
For finite temperature we obtain
\begin{eqnarray}
        \gamma_{collapse}
      =
        0.876.. \frac{k_BT}{\hbar}N^{-1/3}
        \Big(\frac{k_BT_ca^2m}{\hbar^2}\Big)^{1/2}
\end{eqnarray}
By comparison with (\ref{gamma0}) we see that $\gamma_{collapse}$ is of the order of $\gamma_0$ and is therefore in large condensates smaller than $\gamma_c$. 

In summary, the phase-collapse is not effective in large condensates because it occurs with a rate $\gamma_{collapse} < \gamma_c$ and is at the same time restricted to a time interval $\Delta t<\frac{1}{\gamma_c}$ since for larger times phase-diffusion takes over.

\subsection{Phase-diffusion rate}

The phase-diffusion coefficient is a somewhat complicated quantity because it receives contributions from several processes, which are physically distinct. We consider the different contributions separately and also distinguish the two temperature regimes of high temperature $k_BT > \mu$, for which we give the result first, and $k_BT < \mu$ (low temperature).

\subsubsection{Low frequency condensate number fluctuations}

From eq.(\ref{eq:PD}) we infer with $\Gamma_0=\Gamma_{00}$
\begin{eqnarray}
       D^{(\alpha)}_{\phi}=\frac{1}{\hbar k_BT\langle N_0\rangle\Gamma_{00}}
              \left(\langle\Delta N_0^2\rangle
                \frac{\partial\mu}{\partial\langle N_0\rangle}\right)^2
\end{eqnarray}
which is evaluated as
\begin{eqnarray}
       D^{(\alpha)}_{\phi}= 0.0853..\frac{T}{T_c}
        \left(\frac{k_BT_cm a^2}{\hbar^2}\right)^{1/4}
              \frac{1}{\langle N_0\rangle^{1/2}N^{1/6}}\,\,              \frac{k_BT_c}{\hbar}.\label{Dalpha}
\end{eqnarray}
The same result holds in the low temperature regime $k_BT<\mu$. In comparison with $\gamma_0$ (\ref{gamma0}) it is of the order
\begin{eqnarray}
       D^{(\alpha)}_{\phi} \sim N^{-1/3}
           \left(\frac{k_BT_cm a^2}{\hbar^2}\right)^{-1/4}
                 \gamma_0 \sim \sqrt{\frac{\hbar\omega_0}{\mu}}\,\,\gamma_0
\end{eqnarray}
and is much smaller in large and strongly interacting condensates.
Still this contribution to the phase-diffusion rate is always the dominant one
at low temperatures and may dominate even at higher temperatures (see below).

\subsubsection{Condensate number fluctuations due to quasi-particles around energies $\mu$}

Splitting $\Gamma_0 = \Gamma_{00}+(\Gamma_0-\Gamma_{00})$ and expanding to first order
\begin{eqnarray}
       \frac{1}{\Gamma_0}
     = \frac{1}{\Gamma_{00}} - \frac{\Gamma_0 -\Gamma_{00}}{\Gamma_{00}^2}
\end{eqnarray}
we estimate as contribution $D^{(\beta)}_{\phi}$ from the higher frequency condensate number fluctuations described by $\Gamma_0 -\Gamma_{00}$ as given by (\ref{Diff})
\begin{eqnarray}
       D^{(\beta)}_{\phi} \sim - \frac{T}{T_c}\frac{1}{\langle N_0\rangle}
                 \frac{k_B T_c}{\hbar}
\end{eqnarray}
which is in absolute value smaller than the contribution $D^{(\alpha)}_{\phi}$ from low-energy excitations by the order of magnitude factor $\sqrt{\hbar\omega_0/\mu}$. This contribution is therefore negligible in very large condensates. In not so large condensates the complete integral in the result for $\Gamma_0$ needs to be evaluated.

In the low-temperature regime $k_BT<\mu$ we get instead
\begin{eqnarray}
      D^{(\beta)}_{\phi} = -0.306..\frac{1}{N}
              \left(\frac{k_BT_c a^2m}{\hbar}\right)^{1/2}
                    \frac{k_B T_c}{\hbar}
\end{eqnarray}
This is much smaller than $D^{(\alpha)}_{\phi}$ by the order of magnitude factor $(\mu/k_BT_c)^{1/2} / N^{1/3}$.

\subsubsection {Fluctuations in the thermal cloud at energies of order $\mu$}

By eq.(\ref{eq:PD}) this contribution is given by
\begin{eqnarray}
       D^{(\gamma)}_{\phi} = \frac{k_BT}{\hbar\langle N_0\rangle}
              (\Gamma_0 + \Gamma^{\prime})
\end{eqnarray}
which is evaluated to
\begin{eqnarray}
       D^{(\gamma)}_{\phi} = 0.0243..B'^{(\mu)}
              \left(\frac{T}{T_c}\right)^3
              \frac{N}{\langle N_0\rangle^2}
         \left(\frac{k_BT_ca^2m}{\hbar^2}\right)^{\frac{1}{2}}
                    \frac{k_BT_c}{\hbar}.\label{Dgamma}
\end{eqnarray}
This contribution  differs from $D^{(\alpha)}_{\phi}$ by the order of magnitude factor
$(T/T_c)^2\sqrt{\mu\hbar\omega_0}/k_BT_c$ and is therefore much smaller.

For temperatures $k_BT<\mu$ we find instead
\begin{eqnarray}
       D^{(\gamma)}_{\phi} = 0.366..
           \left(\frac{T}{T_c}\right)^8
            \frac{1}{\langle N_0\rangle}\left(\frac{k_BT_ca^2m}{\hbar^2}\right)^{-2}\frac{k_BT_c}{\hbar}
\end{eqnarray}
which is again negligibly small compared to $D^{(\alpha)}_{\phi}$.

In summary,  the phase-diffusion is caused dominantly by the low-frequency particle number fluctuations in the condensate and the phase-diffusion constant is given by eq.(\ref{Dalpha}).
It is proportional to temperature, and scales proportional to $N^{-2/3}$ for fixed $T_c$ or proportional to $N^{-1/2}$ for fixed volume of the trap.

\section{Evaluation of the transport parameters for an isotropic
harmonic trap}\label{sec:7}

In this section we consider the more realistic case of condensates in 
a parabolic trapping potential $m\omega_0^2x^2/2$, which we assume to be isotropic for simplicity. 
In order to analyse the noise $\Im(\xi(t)$ driving
 the fluctuations of $|\alpha_0|^2$ we must  
 consider in detail the relevant linear combination of matrix elements
\begin{equation}
(M^{(1)}_{\kappa,\nu\mu})^*
         -M^{(2)}_{\nu\mu,\kappa} = 2U_0\int d^3 x \psi_0  
             \left\{  (\tilde u_\kappa- \tilde v_\kappa) (\tilde u^*_\mu \tilde v^*_\nu + \tilde v^*_\mu 
                      \tilde u^*_\nu) 
                      -  \tilde u_\kappa \tilde u^*_\mu \tilde u^*_\nu +
                           \tilde v_\kappa \tilde v^*_\mu \tilde v^*_\nu             \right\}
\label{eq:M}
\end{equation}

In the following we shall make use of the local density and Thomas-Fermi approximation, restricting ourselves to large condensates. For the high-lying states we can then use the local energies in Thomas-Fermi approximation
\begin{equation}
       E(p,\bbox x) = \sqrt{(\frac{p^2}{2m}
                  + |U_0n_0(\bbox x)|)^2-U_0^2n_0^2(\bbox x)\Theta(\mu-V(\bbox x))}                       
\end{equation}
with the condensate density
\begin{equation}
n_0(\bbox x)=\langle N_0\rangle|\tilde{\psi}_0(\bbox x)|^2=(\langle\mu\rangle/U_0)(1- (x/r_{TF})^2)
\end{equation}
and the  Thomas-Fermi radius
 \begin{equation}
r_{TF}=\sqrt{2\langle\mu\rangle/m\omega_0^2}=(\frac{15U_0\langle N_0\rangle}{8\pi\langle\mu\rangle})^{1/3}.
\end{equation}
The high-lying quasi-particle modes can be represented similarly to the spatially homogeneous case as
\begin{eqnarray}
u_\kappa(\bbox x)= && \frac{E_\kappa+p^2_\kappa/2m}{\sqrt{2E_\kappa p^2_\kappa/m}}e^{i\bbox{p}_\kappa\cdot\bbox x/\hbar}\nonumber\\
\label{eq:N}\\
v_\kappa (\bbox x)= && - \frac{E_\kappa-p^2_\kappa/2m}{\sqrt{2E_\kappa p^2_\kappa/m}}e^{i\bbox{p}_\kappa\cdot\bbox x/\hbar}\nonumber
\end{eqnarray}
The low-lying collective modes can be represented as
\begin{eqnarray}
 u_\nu (\bbox x)= && \left( \sqrt\frac{U_0 n_0 (\bbox x)}{2\hbar \tilde\omega_\nu}+ 
 \frac{1}{2}\sqrt\frac{\hbar\tilde\omega_\nu}{2U_0n_0(\bbox x)}\right)
\chi_\nu(\bbox x)\nonumber\\
 \label{eq:L}\\
v_\nu(\bbox x)= 
 && \left(-\sqrt\frac{U_0n_0(\bbox x)} {2\hbar\tilde\omega_\nu}+ 
 \frac{1}{2} \sqrt\frac{\hbar\tilde\omega_\nu}{2U_0n_0(\bbox x)}\right)\chi_\nu (\bbox x)\nonumber
\end{eqnarray}
with
\begin{equation}
       \int\!\!d^3 x | \chi_\nu(\bbox x) |^2 = 1
\end{equation}
The mode-functions  $\chi_\nu (\bbox x)$ are known in the hydrodynamic (long-wavelength) and Thomas-Fermi approximation \cite{stringari,fetter,ob,fliesser,csordas} by analytic solutions of the Bogoliubov-equations. In spatially isotropic parabolic traps they have the form 
\cite{stringari}
\begin{equation}
\chi_\nu(\bbox x)=\frac{1}{r_{TF}^{3/2}}P_{\ell_\nu}^{(2n_\nu)}(x/r_{TF})(x/r_{TF})^\ell Y_{\ell m}(\theta,\varphi)\Theta(1-x/r_{TF})
\end{equation}
The polynomials $P_\ell^{(2n)}(x)$ of degree $2n$ are the normalized solutions of the radial part of the Bogoliubov-Fetter equations in the Thomas-Fermi and long-wavelength limit \cite{stringari,fetter}
given by \cite{fetter}
\begin{equation}
P_\ell^{(2n)}(x)=\frac{\sqrt{4n+2\ell+3}}{n!}x^{-2\ell-1}\frac{d^n}{d(x^2)^n}\big [x^{2n+2\ell+1}(1-x^2)^n\big ]
\end{equation}
with the normalization
\begin{equation}
\int_0^1dxx^{2\ell+2}\big [P_\ell^{(2n)}(x)\big ]^2=1
\end{equation}

In the phonon part of the excitation spectrum we have $u_\lambda\simeq
-v_\lambda\sim\tilde\omega^{-1/2}_\lambda$. Furthermore, in that low- energy region
the statistical factor in eqs.(\ref{34}), (\ref{31}), (\ref{31a}) is well approximated by 
$\bar{n}_\nu\bar{n}_\mu\bar{n}_\kappa\approx (k_BT)^3/\hbar^3\tilde\omega_\kappa\tilde\omega_\nu\tilde\omega_\mu$. Just as in the case of box-like traps the frequency
factors in the denominator, together with similar further factors
in the denominator coming from the matrix elements,
make the phonon contribution to the sums in
(\ref{34}) the dominant one, at least in large condensates, and we shall
therefore concentrate on this contribution in the following. This frequency range has a natural upper cut-off at $\langle\mu\rangle/\hbar$, where
the collective phonons go over smoothly into particle-like
excitations. 

For $E_\kappa,E_\nu,E_\mu\ll\langle\mu\rangle$ the matrix elements $(M^{(1)}_{\kappa,\nu\mu})^*,  M^{(2)}_{\nu \mu,\kappa}$ are given by    the integral
\begin{eqnarray}
(M^{(1)}_{\kappa,\nu\mu})^*&\approx& -M^{(2)}_{\nu \mu,\kappa}\label{M}\\
&\approx&-\sqrt{\frac{15}{8\pi}}\frac{3U_0\langle\mu\rangle^{3/2}\delta_{m_\kappa,m_\nu+m_\mu}}{r_{TF}^3\sqrt{2E_\nu E_\mu(E_\nu+E_\mu)}}
J(n_\kappa,n_\nu,n_\mu;\ell_\kappa,\ell_\nu,\ell_\mu)C(\ell_\kappa|\ell_\mu m_\mu,\ell_\nu m_\nu)\label{me}
\end{eqnarray}
where $J$ denotes the integral
\[
J(n_\kappa,n_\nu,n_\mu;\ell_\kappa,\ell_\nu,\ell_\mu)=\int_0^1dxx^2(1-x^2)^2x^{\ell_\kappa+\ell_\nu+\ell_\mu}P_{\ell_\kappa}^{(2n_\kappa)}(x)P_{\ell_\nu}^{(2n_\nu)}(x)P_{\ell_\mu}^{(2n_\mu)}(x)
\]
and the Clebsch-Gordon coefficients $ C(\ell_\kappa|\ell_\mu m_\mu,\ell_\nu m_\nu)$ are given by the angle-integral
\[
C(\ell_\kappa|\ell_\mu m_\mu,\ell_\nu m_\nu)=\int d\Omega Y^*_{\ell_\kappa,m_\nu+m_\mu}(\theta,\varphi)Y_{\ell_\mu,m_\mu}(\theta,\varphi)Y_{\ell_\nu,m_\nu}(\theta,\varphi)
\]
if $|\ell_\mu-\ell_\nu|\leq\ell_\kappa\leq\ell_\mu+\ell_\nu $, otherwise they vanish.
Later-on we shall have to calculate e.g.
$\sum_{m_\nu,m_\mu}|(M^{(1)}_{\kappa,\nu\mu})^* - M^{(2)}_{\nu \mu,\kappa}|^2$
where we can make use of the sum-rule, for 
$|\ell_\mu-\ell_\nu|\leq\ell_\kappa\leq\ell_\mu+\ell_\nu $
\[
\sum_{m_\nu,m_\mu}|C(\ell_\kappa|\ell_\mu m_\mu,\ell_\nu m_\nu)|^2=1
\]
so that the Clebsch-Gordon coefficients need actually not be used explicitly.

In order to have well-defined expressions for the rate-coefficients we again need to smoothen the delta-function expressing energy-conservation, which is done physically by experimental imperfections or limitations in resolution. 
Here this can be done by
replacing the discrete sum over the 'quantum-number' $\ell_\kappa$ by an integral
\begin{eqnarray}
\sum_{\ell_\kappa}\delta(E_\kappa-E_\nu-E_\mu)({\bf ...})&\approx&\int dE_\kappa\frac{1}{dE_\kappa/d\ell_\kappa}\delta(E_\kappa-E_\nu-E_\mu)({\bf ...})\nonumber\\
&=&\int dE_\kappa\frac{E_\nu+E\mu}{(\hbar\omega_0)^2(n_\kappa+1/2)}\delta(E_\kappa-E_\nu-E_\mu)({\bf ...})
\end{eqnarray}
where we used the expression for the excitation-energies \cite{stringari}
\begin{eqnarray}
E_\kappa&=&\hbar\omega_0e_\kappa\nonumber\\
e_\kappa&=&\sqrt{2n_\kappa^2+2n_\kappa\ell_\kappa+3n_\kappa+\ell_\kappa}.
\end{eqnarray}
We introduced the dimensionless eigenvalues $e_{\kappa,\nu,\mu}$
which will appear in the ensuing expressions from now on.
The integration over $E_\kappa$ with the delta-function then picks out the energy-value $E_\kappa=E_\nu+E_\mu$ 
so that $\ell_\kappa$ becomes a function $\ell_\kappa^{(0)}$ of the other 'quantum-numbers'
\[
\ell_\kappa^{(0)}=\frac{(e_\nu+e_\mu)^2-2n_\kappa^2-3n_\kappa}{2n_\kappa+1}
\]
The inequalities
\[
|\ell_\nu-\ell_\mu|\leq\ell_\kappa^{(0)}\leq\ell_\nu+\ell_\mu
\]
then imply that $n_\kappa$ must lie in the interval
\[
n_{\kappa -}\leq n_\kappa\leq n_{\kappa +}
\]
with
\[
n_{\kappa\pm}=\frac{1}{2}\big (\sqrt{|\ell_\nu\pm\ell_\mu|^2+|\ell_\nu\pm\ell_\mu|+9/4+2(e_\nu+e_\mu)^2}-|\ell_\nu\pm\ell_\mu|-3/2\big )
\]

Using all this we obtain from eq.(\ref{34})
\begin{eqnarray}
\Gamma_0=B_{00}\big (\frac{a}{d_0}\frac{k_BT}{\hbar\omega_0}\big )^2
=B_{00}\frac{T^2}{T_c^2}\frac{k_BT_ca^2m}{\hbar^2}\big (\frac{N}{\zeta(3)}\big )^{1/3}\label{res1}
\end{eqnarray}
where $\hbar\omega_0$ is now eliminated in favour of $k_BT_c$ via the relation $\hbar\omega_0=k_BT_c(\zeta(3)/N)^{1/3}$, and where the temperature- and particle-number-independent positive real 
number $B_{00}$ is defined 
by the multiple sums
\begin{equation}
B_{00}=\frac{135\pi^2}{2}\sum_{n_\nu}\sum_{n_\mu}
\sum_{\ell_\nu}\sum_{\ell_\mu}\sum_{n_\kappa=n_{\kappa -}}^{n_{\kappa +}}\frac{(2\ell_\kappa^{(0)}+1)J^2(n_\kappa,n_\nu,n_\mu;\ell_\kappa^{(0)},\ell_\nu,\ell_\mu)}{e_\nu^2e_\mu^2(e_\nu+e_\mu)(2n_\kappa+1)}
\end{equation}
The result for $\Gamma_0$ agrees, except for the numerical prefactor, with the result of
\cite{PRL} which was evaluated there using the local-density approximation
and imposing a lower cut-off for the excitation-frequencies at the geometrical mean trap-frequency $\bar{\omega}$.
It can also be compared with the corresponding result (\ref{G00}) for the box-like trap, which shows the same dependence on temperature and particle-number (if we stipulate $\langle N_0\rangle\sim N$), but the comparison of the prefactor  is problematic because the condensate in the parabolic trap has two length-scales $d_0$ and $r_{TF}$, whereas in the box-like trap only the length-scale L is relevant.  

The property (\ref{M}) of the matrix-elements implies that
the {\it low-lying} excitations don't contribute to $\Gamma_0+\Gamma'$. The reality of the matrix-elements furthermore implies that $\Gamma''$ vanishes.
These remarkably simple results mean that the noise-source $\xi(t)$
introduced in eq.(\ref{eq:1})
is purely imaginary, corresponding to total squeezing in the direction
of the phase $\phi$. In other words, the coupling of the condensate to the 
collective excitations introduces a {\it direct} Langevin noise-source only  for
the number-fluctuations $\delta N_0$, not the phase-variable $\phi$.\footnote{The latter is of course affected by the noise-source  indirectly, 
because the fluctuations of $\delta N_0$ driven by the latter
cause fluctuations 
in the chemical potential.} 

The fact that $\Gamma_0+\Gamma'=0$ for the contribution from the low-lying states implies that the contributions from the higher lying states
must also be considered in order to evaluate the small (compared to $\Gamma_0$)
but finite value
of this quantity. For this purpose we need to consider the matrix-element 
$(M'^{(1)}_{\kappa,\nu\mu})^*
+M'^{(2)}_{\nu\mu,\kappa}$. It differs from the matrix-elements we have considered so far by the replacement $\tilde\psi_0(x)\rightarrow\tilde\psi_0(x)+2\langle N_0\rangle\partial\tilde\psi_0(x)/\partial\langle N_0\rangle$ in the
matrix-element. In Thomas-Fermi approximation this is tantamount to the replacement 
\begin{equation}\tilde\psi_0(x)\rightarrow (2/5)\tilde\psi_0(x)/(1-x^2/r_{TF}^2).\label{wei}
\end{equation}

Physically this implies a  reduced coupling of the thermal fluctuations with the center of the condensate and a strongly enhanced coupling at its boundary, as one would expect for fluctuations located in the thermal cloud.
A mathematical consequence is the fact that the integrals defining these matrix-elements diverge at the boundary in the Thomas-Fermi approximation,
meaning that we encounter here the limitations of that approximation.
Instead of a full-fledged extension of the theory beyond the Thomas-Fermi approximation it will be sufficient for our purposes here to cure its deficiencies by substituting as a cut-off the finite thickness of the boundary-layer given by \cite{rev}
\[d=\frac{1}{2}r_{TF}\left(\frac{\hbar\omega_0}{\langle\mu\rangle}\right)^{2/3}
\]
The matrix-element itself is then evaluated in the local-density approximation \cite{rev}, where we can make use to good purpose of the analysis already performed in the predeeding section. The finte volume $V=L^3$ (and the
associated $\hbar\omega_0=(2\pi\hbar)^2/2mL^2$ which is not to be confused with the trap frequency called $\omega_0$ in the present section) is then an arbitrary local subvolume of the condensate, introduced merely as a technical device like a quantization-volume. It must be sufficiently small so that the condensate within it can be treated as homogeneous, and sufficiently large that we can replace sums over local momenta by integrals. At the end we have to check for consistency whether the result is indeed independent of the choice of
this volume. The result obtained in this way is the local average of the result (\ref{endl}) for the homogneous case, which now becomes space-dependent, because we have to substitute a space-dependent chemical potential $\mu\rightarrow\langle\mu\rangle (1-x^2/r_{TF}^2)$.
This local result can be written as 
\[
\Gamma_0(x)+\Gamma'(x)=\frac{B'^{(\mu)}}
{2\pi^2}\frac{(k_BT)^2a^2m}{\hbar^2\langle\mu\rangle(1-x^2/r_{TF}^2)}
\]
and is indeed independent of the choice of $V$.
The local average has to be performed with the weight $(\tilde\psi_0(x)+2\langle N_0\rangle\partial\tilde\psi_0(x)/\partial\langle N_0\rangle)^2$ determined from (\ref{wei}).
Doing the average and regulating the divergency of the integral at the boundary of the condensate by the physical cut-off we obtain
\begin{equation}
\Gamma_0+\Gamma'=\frac{3}{10}\frac{2^{1/3}}{15^{2/15}\pi^2}B'^{(\mu)}
\left(\frac{k_BT}{\hbar\omega_0}\right)^2\langle N_0\rangle^{-2/15}\left(\frac{a}{d_0}\right)^{28/15}=0.024..B'^{(\mu)}\left(\frac{T}{T_c}\right)^2\left(\frac{N}{\langle N_0\rangle}\right)^{2/15}N^{2/9}
\left(\frac{k_BT_ca^2m}{\hbar^2}\right)^{14/15}.\label{sch}\end{equation}

In order to extract results for the relaxation-rate of the condensate number
and the phase-diffusion rate it is necessary to know also the 
mean square of the number-fluctuations $\langle \Delta N_0^2\rangle$.
This can be evaluated from eq.(\ref{deltan}),  using the fact that these fluctuations are
also dominated by the low-lying modes \cite{11a}.
The result of this calculation to leading order in $(\hbar\omega_0/k_BT)$
is
\begin{eqnarray}
\langle\Delta N_0^2\rangle=&&A\big(\frac{\langle N_0\rangle a}{d_0}\big)^{4/5}\big(\frac{k_BT}{\hbar\omega_0}\big)^2\label{Dela}\\
=&&\frac{A}{(\zeta(3))^{8/15}}\left(\frac{T}{T_c}\right)^2\left(\frac{\langle N_0\rangle}{N}\right)^{4/5}N^{4/3}\left(\frac{k_BT_ca^2m}
{\hbar^2}\right)^{2/5}
\label{Delta}\end{eqnarray}
with the number $A$ given by the multiple sums
\begin{equation}
A=\frac{(15)^{4/5}}{2}\sum_n\sum_{n'}\sum_\ell
\frac{2\ell+1}{(e(n,\ell)e(n',\ell))^2}\left|\int_0^1 dx(1-x^2)x^{2(\ell+1)}P_\ell^{(2n)}(x)P_\ell^{(2n')}(x)\right|^2
\end{equation}
In order to find the scaling of $\langle \Delta N_0^2\rangle$ in the thermodynamic limit $N\rightarrow\infty, \omega_0\rightarrow 0, k_BT_c=\hbar\omega_0(N/\zeta(3))^{1/3} fixed$ it is necessary to use the form
of the preceding results in which
$\hbar\omega_0$ is eliminated in favor of $k_BT_c$ and to use  $\langle N_0\rangle\sim N$.. Then the scaling
$\langle \Delta N_0^2\rangle\sim N^{4/3}$ derived in \cite{11a} is
 recovered.
The particle-number relaxation rate now follows from eqs.(\ref{tau}) and (\ref{res1}) as
\begin{equation}
\gamma_c=\frac{2B_{00}}{A}\langle N_0\rangle^{1/5}\left(\frac{a}{d_0}\right)^{6/5}\frac{k_BT}{\hbar}=\frac{2(\zeta(3))^{1/5}B_{00}}{A}\frac{T}{T_c}\left(\frac{\langle N_0\rangle}{N}\right)^{1/5}\left(\frac{k_BT_ca^2m}{\hbar^2}\right)^{3/5}\frac{k_BT_c}{\hbar}
\end{equation}
It is the largest of the various rates we calculate here but is still small compared to $\omega_0$, the inverse time-scale of motion in the trap, by the order of magnitude $N^{-2/3}(Na/d_0)^{6/5}$.

The phase-collapse rate is obtained from (\ref{coll}).
At $T\ne 0$ ( more precisely above a cross-over temperature of order $\hbar\omega_0$) we find
\begin{equation}
\gamma_{collapse}=\frac{15^{2/5}A^{1/2}}{5}\langle N_0\rangle^{-1/5}\left(\frac{a}{d_0}\right)^{4/5}\frac{k_BT}{\hbar}=\frac{15^{2/5}(\zeta(3))^{2/15}A^{1/2}}{5}\frac{T}{T_c}\left(\frac{N}{\langle N_0\rangle}\right)^{1/5}N^{-1/3}\left(\frac{k_BT_ca^2m}{\hbar^2}\right)^{2/5}\frac{k_BT_c}{\hbar}.
\end{equation}
Apart from the numerical prefactor this is the same asymptotic expression as obtained for the damping-rate $\gamma_0$ of the low-lying collective modes (see e.g. \cite{Nic}). It is smaller than $\gamma_c$ by the order of magnitude
$(\langle N_0\rangle a/d_0)^{-2/5}$, i.e. the phase-collapse remains inefficient before phase-diffusion takes over.

The phase-diffusion constant $D_\phi^{(\alpha)}$ due to the exchange of particles between the condensate and low-lying excitations is gotten by inserting the results for $\langle\Delta N_0^2\rangle$ and $\Gamma_{00}$ in eq.(\ref{eq:PD}):
\begin{equation}
D_\phi^{(\alpha)}=\frac{(15)^{4/5}A^2}{25B_{00}}\langle N_0\rangle^{-3/5}\left(\frac{a}{d_0}\right)^{2/5}\frac{k_BT}{\hbar}=\frac{15^{4/5}(\zeta(3))^{1/15}A^2}{25B_{00}}\frac{T}{T_c}\left(\frac{N}{\langle N_0\rangle}\right)^{3/5}N^{-2/3}\left(\frac{k_BT_ca^2m}{\hbar^2}\right)^{1/5}\frac{k_BT_c}{\hbar}.
\end{equation}
It is smaller than $\gamma_{collapse}$, again by the order of magnitude of $(\langle N_0\rangle a/d_0)^{-2/5}$. 

Finally, the contribution of the fluctuations in the thermal cloud to the phase-diffusion is also obtained from (\ref{eq:PD}) by inserting the result (\ref{sch})
for $\Gamma_0+\Gamma'$:
\begin{eqnarray}
D_\phi^{(\gamma)}&&=\frac{3}{10}\frac{ 2^{1/3}B'^{(\mu)}}{ 15^{2/15}\pi^2}\langle N_0\rangle^{-17/15}\left(\frac{a}{d_0}\right)^{28/15}\left(\frac{k_BT}{\hbar\omega_0}\right)^2\frac{k_BT}{\hbar}\nonumber\\
&&=\frac{3}{10}\frac{ 2^{1/3}(\zeta(3))^{-16/45}B'^{(\mu)}}{ 15^{2/15}\pi^2}\left(\frac{T}{T_c}\right)^3\left(\frac{N}{\langle N_0\rangle}\right)^{17/15}N^{-7/9}\left(\frac{k_BT_ca^2m}{\hbar^2}\right)^{14/15}\frac{k_BT_c}{\hbar}.
\end{eqnarray}
It differs from the previous rates, which were all proportional to temperature, by the stronger temperature-dependence $\sim T^3$. However, this contribution to $D_\phi$ remains smaller than $D_\phi^{(\alpha)}$ by the order of magnitude $N^{-1/9}(k_BT_ca^2m/\hbar^2)^{11/15}(T/T_c)^2$.

\section{Discussion and Conclusion}
\label{sec:}

In this paper we have put forward a detailed theory  of fluctuations and relaxation processes of the condensate in thermal equilibrium with the cloud of its excitations. For a given number of particles $N_0$ in the condensate, we have defined the condensate mode as the corresponding normalized solution of the Gross-Pitaevskii equation, defining at the same stroke the $N_0$-dependent part of the chemical potential. The equilibrium value of $\langle N_0\rangle$ is distinguished as the value of $N_0$ for which the number of particles in the thermal cloud {\it in equilibrium} with the condensate plus $N_0$ equals $N$. We have calculated the fluctuations of $N_0$ around its equilibrium value and also the fluctuations of the phase of the complex amplitude $\alpha_0$ of the condensate with $|\alpha_0|^2=N_0$. In a general phenomenological framework presented in the first part of this paper we were able to separate the fluctuations of the complex condensate amplitude into several contributions, which have different physical origin:

-- The fluctuation of the atom-number in the condensate, which are driven by the exchange of atoms between the condensate and the thermal cloud.

-- The fluctuation of the chemical potential with two different contributions, namely the fluctuations of $\mu$ due to number-fluctuations in the condensate, and the faster fluctuations of $\mu$ at constant $N_0$ caused by number-fluctuations in the excitations.

The importance of number-fluctuations in the condensate, assumed at first in the phenomenological approach due to the importance of $N_0$ for the value of the chemical potential, but later born out by the microscopic calculations, leads to the appearance of the linear relaxation-rates $\gamma_c$ of the condensate-number as an important characteristic inverse time-scale of the problem. At times much shorter than $\gamma_c^{-1}$ phase-diffusion of the condensate-phase due to the fast number-fluctuations in the excitations can occur. In the same regime may also occur the process of collapse due to the reversible spreading of the phase caused by the static uncertainty in $N_0$ and the associated chemical potential.

At times large compared to $\gamma_c^{-1}$ the number-fluctuations in the condensate are dynamical and irreversible, and lead to the replacement of the reversible collapse by an irreversible phase-diffusion with a larger diffusion-rate than in the short-time regime.

The second and larger part of this paper was devoted to microscopic theory. First we have provided a microscopic derivation of the phenomenological Langevin equation, established microscopic formulas for all phenomenological parameters and also exhibited the relation between the short-time diffusion rate and fluctuation rates of the population numbers of excitations via a sum-rule. Then the microscopic theory was used to evaluate the transport-parameters and the various rates as a function of temperature, particle-number and the scattering length of the interaction potential. The evaluation was done for two simple cases -- the cubic box-like trap, where the form of the condensate mode does not depend on $N_0$ and the thermal cloud penetrates the condensate homogeneously, and the isotropic harmonic trap, where the form of the condensate-mode changes with $N_0$ and the thermal cloud is located preferentially near the boundaries of the condensate. The physically important results for both kinds of traps are similar, even though they have to differ, obviously, in the details of the scalings with the atom-numbers and the scattering length.

The calculation of the transport-parameters reveals some interesting physical results:

-- The fluctuations driving the absolute value $|\alpha_0|$ and the phase $\phi$ of $\alpha_0$ are quite different in strength, those driving $|\alpha_0|$ being the much stronger ones. The reason for this is a pronounced squeezing of the bath of thermal excitations with respect to the instantaneous phase of the condensate. This squeezing reaches nearly 100\%~for the lowest-lying modes, which is the reason that fluctuations of $\phi$ are practically not driven by such modes. On the other hand, the contribution of the high-lying modes to the fluctuating forces driving $|\alpha_0|$ and $\phi$ is nearly the same (after the obvious normalization with $|\alpha_0|$), i.e. there is no squeezing in this (much weaker) contribution to the noise.

-- The cross-correlation between the fluctuations driving $|\alpha_0|$ and $\phi$ are found to vanish exactly in a real condensate, where both the Gross-Pitaevskii equation and the Bogoliubov-Fetter equations are real and all solutions can  (but need not) be taken real. This can also be understood as a general consequence of time-reversal symmetry: $\phi$ is a velocity potential and therefore odd under time-reversal while $|\alpha_0|$ is even under time-reversal. Their fluctuating forces therefore transform oppositely. In a time-reversal symmetric condensate (no vortices) the cross-correlation between an even and an odd quantity under time-reversal must vanish.

It turns out that the relaxation rate $\gamma_c$ of the atom-number in the condensate is the largest of the calculated rates. In particular it is larger than the collapse-rate and the  phase-diffusion rate which, like $\gamma_c$, are proportional to temperature in the regime $k_BT>\mu$. It is also larger than the decay rates of the lowest-lying collective modes $\gamma_0$, which might
look surprising because at the same time the theory tells us that $\gamma_c$ is dominated by the particle transfer-rates between the condensate and the low-lying modes. However, it is clear that $\gamma_c$ ought to be larger than $\gamma_0$ because the condensate couples to all low-lying modes in parallel which increases the number of decay channels by a factor proportional
to the ratio of the chemical potential and the lowest-lying mode frequency.

The next largest rate we find is the thermal phase-collapse rate $\gamma_{collapse}$. It turns out to have the same functional dependence on $T, a, \langle N_0\rangle$ and $N$ as the decay-rate of the lowest-lying collective modes. I cannot see any fundamental reason for this coincidence and have to count it just as that. Physically the smallness of $\gamma_{collapse} / \gamma_c$ means that the phase-collapse will not be observable at finite temperature because it can only lead to a decay-factor $exp(-\frac{1}{2}(\gamma_{collapse}/\gamma_c)^2)$ very close to 1 before phase-diffusion takes over.

Finally, the phase-diffusion rate $D_{\phi}$ is the smallest of the rates calculated here. We find the simple nice result that the ratios $\gamma_c/\gamma_0$ and $\gamma_0/D_{\phi}$ are of about equal order of magnitude, given by the ratio of $\mu$ to the smallest excitation energy, which is $\hbar\omega_0$ for the harmonic and $\sqrt{\mu/m}(2\pi\hbar/L)$ for the box-like trap. Instead of $\gamma_0$ we may also take $\gamma_c$ in these ratios with the same conclusion. $D_{\phi}$ like the rate $\gamma_c$ is found to be dominated by the atom-number exchange between the condensate and the low-lying modes. 

This observation actually explains the coincidence of the two ratios we have just indicated and turns them into a precise relation:
In (\ref{eq:PD}) for $D_{\phi}$ we put $\Gamma^"=0$ which is exact for real condensate-modes and neglect $\Gamma_0+\Gamma^{\prime}$, which comes from high-lying excitations. Then multiplying the resulting expression for $D_{\phi}$, with $\gamma_c=\tau_c^{-1}$ from (\ref{tau}), we readily find
\begin{equation}
       \frac{1}{2}D_{\phi}\cdot\gamma_c 
      = \gamma_{collapse}^2 
\end{equation}
with $\gamma_{collapse}$ from (\ref{coll}) again with $\Gamma^"=0$.

Let us now compare our results with related ones found in the literature. Most closely related to the present work in goal and scope is a paper by Jaksch et al \cite{Jaksch} on the intensity and amplitude fluctuations of a Bose-Einstein condensate at finite temperature, which builds on extensive earlier work by Gardiner and Zoller with collaborators (cf. the references given in \cite{Jaksch}). Unlike the present paper it also takes into account trap losses. The theory presented in \cite{Jaksch} is based on a conceptual division of the Bose gas into two energy regions called the condensate band and the noncondensate band. In this construction the boundary between the two regions is chosen in such a way  that the noncondensate band is not significantly affected by the mean field of the condensate, while the influence of excitations in the condensate band is neglected. Thus the main physical difference of \cite{Jaksch} to the present work is that  it neglects fluctuations of particles from the condensate mode to quasiparticle modes as well as to very low-lying one-particle excitations.

By contrast in the present work we avoid the division of the energy region into two parts. We find, as we have discussed, that the exchange of particles between the condensate and the low-lying modes makes not only an important but in fact the dominant contribution to the relaxation rate of the condensate-number and the phase-diffusion rate, determining their dependence on temperature, atom number, population of the condensate and scattering length. 

The importance of the particle-exchange between the low-lying excitations for the phase-diffusion 
of the condensate and the number-relaxation rate $\gamma_c$ was first pointed out in \cite{PRL}, while for the {\it intensity} of the number-fluctuations in the condensate this had already been shown  in \cite{11a}. The theory
put forward in \cite{PRL} already proceeded along 
essentially the same lines we follow here,
but it  had some short-comings which we overcome and correct
in the present work:
The squeezing of the noise from the thermal cloud with respect to the phase of the condensate was briefly remarked upon in \cite{PRL}, but was not taken
into account in the calculation of the transport coefficients presented there
and in the formula for phase-diffusion. Moreover, in  the conservative part of the Langevin-equation (\ref{eq:1}) $\Delta_0\mu$ was replaced by $\partial\Delta F(|\alpha)|^2)/\partial |\alpha_0|^2$ in \cite{PRL},
which, on scrutiny, appears questionable  when used in conjunction with the
fluctuation formula (\ref{deln},\ref{deltan}) for $\langle\Delta N_0^2\rangle$.
After all, neither $\Delta_0\mu$ nor $\Delta F$ are equilibrium quantities.
The use of the aforementioned relation between them
 is therefore avoided here. 

Even though in the present paper I have opted for the use of the fluctuation-formulas (\ref{deln},\ref{deltan}), which in my opinion have 
a firm basis, it is only fair to mention that
they are still under debate in the current literature, see e.g. \cite{W}.
In another recent paper with some bearing on this topic Bergeman et al. \cite{Ber} use as equilibrium distribution
for the condensate number $P(N_0)\sim\exp[(\langle\mu\rangle N_0-\frac{5}{14}(15 N_0 a/d_0)^{2/5}N_0)/k_BT]$, (cf. the discussion after their eq.(21)), which implies
$\langle \Delta N_0^2\rangle\sim T \langle N_0\rangle^{3/5}$, a result which is rather different, both in the temperature-dependence and in the scaling with the particle-number, from the result (\ref{Dela}) on which our present
calculations have been based.
It is clear that not the method but the details of
our results on the dynamics of the fluctuations
of the condensate would change, if the results on the statics
would be changed. Needless to say that a resolution of the theoretical debate concerning the correct approach to the statics seems urgent and would be highly wellcome. Vice versa experimental results on the dynamics (i.e. on
$\gamma_c$ and $D_\phi$) would also help
to decide, by applying the theory presented here, which of the approaches to the statics of the number-fluctuations in the condensate proposed in the literature
describes the physics correctly.

  A quantum kinetic theory of trapped atomic gases has also been formulated by Stoof \cite{Stoof}. In  \cite{Stoof} the general coupled Fokker-Planck equations of the condensate and the excited modes are presented and applied to the kinetics of the formation of a condensate. This problem has also been studied by Gardiner and coworkers \cite{Ga} as well as Kagan and Svistunov \cite{Ka}, where also earlier work by further authors is quoted.

By contrast the present work has  focussed on the fluctuations around the equilibrium state of the condensate, {\it after} it has been formed.  However, the application of our approach to the kinetics of the formation of a condensate
would be
an interesting goal for future work.

Experimentally the rates $\gamma_c$ and $D_{\phi}$ we have calculated should be measurable. The rate $\gamma_c$ may be observable as the relaxation rate of the condensate back to its equilibrium state after creating a non-equilibrium state by a sudden small change of temperature via evaporative cooling. The sum of the phase-diffusion rates of two condensates could be measured by monitoring the phase-difference between them after it was initially fixed by measurement or preparation at a reference-time $t=0$.
Methods for measuring phase-differences in Bose-Einstein condensates have recently been demonstrated \cite{2,2a,2b}. It is to be hoped therefore that the phase-diffusion in Bose-Einstein condensates - a fundamental process intimately linked to the spontaneously broken gauge symmetry in a finite system - will be measured in the near future.

\section*{Acknowledgement}

Usefull discussions with Walter Strunz are gratefully acknowledged.
This work has been supported by the Deutsche Forschungsgemeinschaft through
the Sonderforschungsbereich 237 ``Unordnung und gro{\ss}e Fluktuationen''.

\appendix\section{}\label{sec:10}

Here we wish to derive the expression (\ref{exp}) for$H_0$.
Using the Gross-Pitaevskii equation (\ref{eq:Ha1}) we put (\ref{H_0}) in the form\begin{eqnarray}
       H_0=(\mu_0-\langle\mu\rangle)|\alpha_0|^2-\frac{U_0}{2}|\alpha_0|^4\int d^3x\psi_0^4
\label{eq:0}
\end{eqnarray}
Taking the derivative with respect  to $|\alpha_0|^2$
we get
\begin{eqnarray}
       \frac{\partial H_0}{\partial|\alpha_0|^2}=\mu_0-\langle\mu\rangle+|\alpha_0|^2\left(\frac{\partial\mu_0}{\partial|\alpha_0|^2}-U_0\int d^3x\psi_0^4-|\alpha_0|^2U_0\int d^3x\psi_0^2\frac{\partial\psi_0^2}{\partial|\alpha_0|^2}\right)\label{+1}
\end{eqnarray}
To evaluate this further we use eq.(\ref{eq:Ha1}) and its derivative with respect to $|\alpha_0|^2$
\begin{eqnarray}
       \left(-\frac{\hbar^2}{2m}\bigtriangledown^2 + V - \mu_0 +
       3U_0|\alpha_0|^2\psi^2_0\right)
       \frac{\partial\psi_0}{\partial|\alpha_0|^2}
       =
       \left(\frac{\partial\mu_0}{\partial|\alpha_0|^2}
       - U_0 \psi^2_0\right)\psi_0
 \label{eq:*}
\end{eqnarray}
Multiplying eq.(\ref{eq:*}) with $\psi_0$ and integrating over space, using the Gross-Pitaevskii equation (\ref{eq:Ha1}) after partial integration, we derive the identity
\begin{eqnarray}
       U_0|\alpha_0|^2 \int d^3x \psi^2_0
       \frac{\partial{\psi}_0^2}{\partial|\alpha_0|^2 } =
       \frac{\partial\mu_0}{\partial|\alpha_0|^2 }-
       U_0 \int\psi^4_0 d^3x
\end{eqnarray}
which is used in  in (\ref{+1}) to yield $\partial H_0/\partial|\alpha_0|^2 =\mu_0-\langle\mu\rangle$, and upon integration results in (\ref{exp}).

\section{}\label{sec:11}

Here we wish to derive eq.(\ref{eq:(G)}). This is achieved if we succeed to show that the coupling of the condensate and the thermal cloud via 
\begin{equation}
\hat H_3=U_0\sqrt{\langle N_0\rangle}\int d^3x\tilde\psi_0\hat{\tilde\chi}^+(e^{-i\phi}\hat{\tilde\chi}+e^{i\phi}\hat{\tilde\chi}^+)\hat{\tilde\chi}\label{B1}
\end{equation}
gives rise to the systematic change of $\Im{(\hat\xi(t))}$, to first order in the interaction, of
\begin{equation}
\delta\langle\Im(\hat\xi(t)\rangle_\phi=-\frac{2\sqrt{\langle N_0\rangle}}{\hbar k_BT}\int_{-\infty}^tdt'S_{JJ}(t-t')\frac{\partial H_0(t')}{\partial|\alpha_0|^2},
\end{equation}
because this can then be used in (\ref{58b}) to yield (\ref{eq:(G)}).
In (\ref{B1}) we could put $|\alpha_0|=\sqrt{\langle N_0\rangle}$ since we linearize around equilibrium and only wish to calculate the dissipation in $|\alpha_0|^2$ which is conjugate to $\phi$, the variable we kept in (\ref{B1}). Standard first order perturbation theory with adiabatic switch-on of the interaction gives, with $\epsilon\rightarrow +0$,
\begin{equation}
\delta\langle\Im(\hat\xi(t)\rangle_\phi=-\frac{i}{\hbar}\int_{-\infty}^tdt'\langle[\Im(\hat\xi(t)),\hat H_3(t')]\rangle_\phi e^{\epsilon t'}.
\label{B3}\end{equation}
We can rewrite this as 
\begin{equation}
\delta\langle\Im(\hat\xi(t))\rangle_\phi=-2i\sqrt{\langle N_0\rangle}\int_{-\infty}^tdt'\left(\chi_{J\hat\xi}''(t,t')e^{-i\phi(t')}+\chi_{J\hat\xi^+}''(t,t')e^{i\phi(t')}\right)e^{\epsilon t'}\label{B4}\end{equation}
where we introduced the response functions
\begin{eqnarray}
\chi_{J\hat\xi}''(t,t')&&=\frac{1}{2\hbar}
\langle[\Im(\hat\xi(t)),\hat\xi(t')]\rangle_\phi\Theta(t-t')\nonumber\\
\label{B5}\\
\chi_{J\hat\xi^+}''(t,t')&&=\frac{1}{2\hbar}
\langle[\Im(\hat\xi(t)),\hat\xi^+(t')]
\rangle_\phi\Theta(t-t')\nonumber
\end{eqnarray}
with
\begin{equation}
\hat\xi(t)=U_0\int d^3x\tilde{\psi}_0\hat{\tilde\chi}^+(t)\hat{\tilde\chi}(t)
\hat{\tilde\chi}(t).\label{B6}\end{equation}
Here $\Theta(t-t')$ is the Heaviside step-function. We shall define $\Theta(0)=0$ without loss of generality.
The fluctuation-dissipation theorem (in the classical frequency domain $\hbar \omega\ll k_BT$) ensures the relations
\begin{eqnarray}
\chi_{J\hat\xi}''(t,t')&&=-\frac{i\Theta(t-t')}{2k_BT}\frac{\partial}{\partial t'}S_{J\hat\xi}(t,t')\nonumber\\
\label{B7}\\
\chi_{J\hat\xi^+}''(t,t')&&=-\frac{i\Theta(t-t')}{2k_BT}\frac{\partial}{\partial t'}S_{J\hat\xi^+}(t,t')\nonumber\end{eqnarray}
with the correlation functions\footnote{$\Im(\hat\xi(t))$ according to eq.(\ref{59b}) contains an explicit external time-dependence via $\phi(t)$, in addition to the internal time-dependence of $\hat{\tilde\chi}(t), \hat{\tilde\chi}^+(t)$ via their Heisenberg equations of motion. This explicit time-dependence has to be taken into account when applying the fluctuation-dissipation theorem. We  avoid this additional step by applying the time-derivative in the fluctuation-dissipation relation (\ref{B7}) directly to the {\it second} time argument $t'$, of course with the appropriate extra minus-sign.}
\begin{eqnarray}
S_{J\hat\xi}(t,t')&&=
\langle\Im(\hat\xi(t))\hat\xi(t')\rangle_\phi\nonumber\\
\label{B8}\\
S_{J\hat\xi^+}(t,t')&&=
\langle\Im(\hat\xi(t))\hat\xi^+(t')
\rangle_\phi.\nonumber
\end{eqnarray}
We can use (\ref{B7}) in (\ref{B4}) and apply a partial integration in $t'$ to obtain
\begin{eqnarray}
\delta\langle\Im(\hat\xi(t))\rangle_\phi=&&\frac{\sqrt{\langle N_0\rangle}}{ik_BT}\int_{-\infty}^tdt'\left(S_{J\hat\xi}(t,t')e^{-i\phi(t')}-S_{J\hat\xi^+}(t,t')e^{i\phi(t')}\right)\frac{d\phi(t')}{dt'}
e^{\epsilon t'}\nonumber\\
&&-\frac{\sqrt{\langle N_0\rangle}}{k_BT}\left(S_{J\hat\xi}(t,t)e^{-i\phi(t)}
+S_{J\hat\xi^+}(t,t)e^{i\phi(t)}\right)\label{B9}
\end{eqnarray}
which can be rewritten as
\begin{eqnarray}
\delta\langle\Im(\hat\xi(t))\rangle_\phi=&&\frac{2\sqrt{\langle N_0\rangle}}{k_BT}\int_{-\infty}^tdt'S_{JJ}(t,t')\frac{d\phi(t')}{dt'}
-\frac{2\sqrt{\langle N_0\rangle}}{k_BT}S_{JR}(t,t)\label{B10}
\end{eqnarray}
with $S_{JJ}(t,t')$ defined by eqs.(\ref{59b},\ref{59c}) and
\begin{equation}
S_{JR}(t,t')=\frac{1}{4i}\langle(\hat\xi(t)e^{-i\phi(t)}-\hat\xi^+(t)
e^{i\phi(t)})(\hat\xi(t')e^{-i\phi(t')}+\hat\xi^+(t')
e^{i\phi(t')})\rangle_\phi.\label{B11}
\end{equation}
The constant term with $S_{JR}(t,t)$ amounts to a small shift of the equilibrium value of $\mu$ in the final result which we shall neglect like other terms contributing to such shifts.
Then using eq.(\ref{59d}) we put $\hbar d\phi(t')/dt'=-\partial H_0(t')/\partial|\alpha_0|^2$ in eq.(\ref{B10}) which establishes (\ref{B1})
and hence eq.(\ref{eq:(G)}).

\end{document}